\documentclass[footinbib, aps,prx,twocolumn,english,floatfix,superscriptaddress,longbibliography]{revtex4-2}

\usepackage{amsmath, amsthm, amsfonts, amssymb,amstext}
\usepackage{mathptmx}
\usepackage{bm}% bold math
\usepackage{xfrac}
\usepackage{mathrsfs}
\usepackage{latexsym}
\usepackage{array}
\usepackage{braket}
\usepackage{amsmath,amssymb,amsfonts}
\usepackage{xcolor}
\usepackage{tikz,tkz-euclide}
\usepackage{dsfont}
\usepackage[colorlinks=true,citecolor=blue,linkcolor=blue,urlcolor=blue]{hyperref}

\def\beq{\begin{equation}}
\def\eeq{\end{equation}}
\def\barray{\begin{eqnarray}}
\def\earray{\end{eqnarray}}

\def\ee{\mathord{\rm e}}

\def\ii{\mathord{\rm i}}
\def\min{\mathord{\rm min}}

\usepackage{graphicx}% Include figure files
\usepackage{dcolumn}% Align table columns on decimal point
\usepackage{bm}% bold math
\begin{document}

\preprint{APS/123-QED}

\title{ Topological crystals and      soliton lattices  in a   Gross-Neveu model   with Hilbert-space  fragmentation }
\author{S. Cerezo-Roquebrún}
\affiliation{Instituto de Fisica Teórica, UAM-CSIC, Universidad Autónoma de Madrid, Cantoblanco, 28049 Madrid, Spain.}%
 \author{S.J. Hands}
 \affiliation{Department of Mathematical Sciences,
 University of Liverpool, Liverpool L693BX, U.K.}%
\author{A. Bermudez}%
\affiliation{Instituto de Fisica Teórica, UAM-CSIC, Universidad Autónoma de Madrid, Cantoblanco, 28049 Madrid, Spain.}%


\begin{abstract}
We explore the finite-density phase diagram of the %(1+1)D
single-flavour Gross-Neveu-Wilson (GNW) model in (1+1) dimensions using matrix product state (MPS) simulations. At zero temperature and along the symmetry line of the phase diagram, we find a sequence of inhomogeneous ground states that arise through a real-space version of the mechanism of Hilbert-space fragmentation. For weak interactions, doping the symmetry-protected topological (SPT) phase of the GNW model leads to localised charges or holes at periodic arrangements of immobile topological defects separating the fragmented subchains: a topological crystal. Increasing the interactions, we observe a transition into a parity-broken phase with a pseudoscalar condensate displaying a modulated periodic pattern. This soliton lattice is a sequence of topological charges corresponding to anti-kinks, which also bind the doped fermions at their respective centres. Out of this symmetry line, we show that quasi-spiral profiles appear with a characteristic wavevector set by the density \( k = 2\pi\rho \), providing non-perturbative evidence for chiral spirals beyond the large-\( N \) limit. These results demonstrate that various exotic inhomogeneous phases can arise in lattice field theories, and motivate the use of quantum simulators to confirm such  QCD-inspired phenomena in future experiments.
\end{abstract}

%keywords{Suggested keywords}%Use showkeys class option if keyword
\maketitle

%\tableofcontents
%%%%%%%%%%%%%%%%%%%%%%%%%%%%%%%%%%%%%
%%%%%%%%%%%%%%%%%%%%%%%%%%%%%%%%%%%%
\setcounter{tocdepth}{2}
\begingroup
\hypersetup{linkcolor=black}
\tableofcontents
\endgroup
%%%%%%%%%%%%%%%%%%%%%%%%%%%%%%%%%%%%
%%%%%%%%%%%%%%%%%%%%%%%%%%%%%%%%%%%%

\section{\textbf{Introduction}}

The phase diagram of  quantum chromodynamics (QCD)~\cite{PhysRev.125.1067,GELLMANN1964214,Fritzsch:1973pi} 
 predicts various exotic phases of matter in which quarks, normally confined into hadrons, become deconfined and form a quark-gluon plasma or  a colour superconductor~\cite{RevModPhys.80.1455}. 
As temperature and chemical potential are lowered, chiral symmetry gets spontaneously broken, leading to a chiral condensate that is responsible for the largest fraction of the observed hadron masses within the confined phase. On the way there,  asymptotic freedom is responsible for the breakdown of perturbation theory~\cite{PhysRevLett.30.1343,PhysRevLett.30.1346}, and it becomes extremely challenging to understand quantitatively how these phases get interconnected via transitions/crossovers,  let alone to find other interesting phases of quark matter. Taking a large-$N$ limit~\cite{HOOFT1974461,0521318270}, where $N$ here stands for the number of quark colours,  it is possible to resum certain leading  Feynman diagrams to all orders of the coupling strength,  allowing for certain non-perturbative predictions of QCD matter. This can lead, for instance, to the suppression of colour superconductivity in favour of a new phase with inhomogeneous wave-like chiral condensates~\cite{doi:10.1142/S0217751X92000302}.
Unfortunately,  subsequent works have estimated that the critical colour number $N>N_{\rm c}$ required to see these so-called chiral waves at moderate chemical potentials lies far from any realistic regime~\cite{SHUSTER2000434}.   More recent studies~\cite{KOJO201037} have argued that inhomogeneous phases could still arise at larger densities, provided that the quarks near the Fermi surface are confined into strongly-interacting baryons ~\cite{MCLERRAN200783}, while those deep inside it contribute to the scaling of the free energy and pressure with $N$~\cite{FUKUSHIMA201399}{; see~\cite{Hands:2010gd} for evidence for this scenario from lattice QCD simulations with the gauge group SU(2)}. These studies identify a specific regime of densities and temperatures for which QCD with a realistic colour number $N_{\rm c}$  can host a chiral spiral: an inhomogeneous phase in which both the scalar and pseudoscalar condensates display a one-dimensional wave-like pattern.
 
A precise confirmation of this inhomogeneous phase at moderate temperatures and densities, as well as a small number of colours, would require a non-perturbative approach such as lattice QCD~\cite{PhysRevD.10.2445,doi:10.1142/6065}. Unfortunately, at finite densities, the standard Monte Carlo sampling techniques used in lattice QCD are compromised due to the sign problem~\cite{NAGATA2022103991,PhysRevLett.94.170201}. Other approaches, such as
complex Langevin~\cite{PhysRevD.81.054508,RevModPhys.94.015006}, Lefshetz thimbles~\cite{PhysRevD.86.074506,refId0}, or tensor network techniques~\cite{PhysRevLett.69.2863,RevModPhys.93.045003}, might develop in the future to the point where they can address finite densities and even real-time phenomena in QCD, {but there remains much work to be done}. Another,  fundamentally different, approach is that of quantum simulators (QSs)~\cite{Feynman1982,Cirac2012}, in which one uses highly-controlled quantum devices to mimic the physics of these models, typically regularised on a spatial lattice. Pioneering works in the context of QFTs, both with global and local symmetries~
\cite{buchler2005atomic,PhysRevLett.98.260402,PhysRevLett.99.201301,PhysRevLett.103.035301,doi:10.1126/science.1217069,PhysRevLett.105.190403,PhysRevLett.105.190404,Weimer2010rydberg,zohar2012simulating,zohar2013cold,banerjee2012atomic,banerjee2013atomic} suggest that QSs may also be able to overcome the limitations of classical numerical approaches as, respectively,  they do not suffer from the aforementioned sign problem, nor from instabilities or convergence issues in the solutions of stochastic differential equations, from the complexity of navigating the complex integration manifold in search for the thimble,  or from the accumulation of entanglement and the inherent complexity of contracting high-dimensional tensors. On the other hand, QSs are subject to environmental noise and experimental control errors in current devices, and big efforts are being devoted to their scaling to the large system sizes that would be required for the exploration of QCD (see the reviews~\cite{Wiese2013ultracold,Zohar2016quantum, dalmonte2016lattice,banuls2020review,Banuls2020simulating,Aidelsburger2021cold,Klco_2022,PRXQuantum.4.027001}).

In view of all this complexity, it is reasonable to start by focusing on models that simplify QCD, a $(3+1)$-dimensional non-Abelian quantum field theory (QFT) but, importantly, share characteristic features with it. In this work, we are interested in asymptotically-free models that can also present
inhomogeneous chiral condensates at finite densities~\cite{BUBALLA201539}. The archetypal example is the Nambu-Jona-Lasinio (NJL) model~\cite{PhysRev.122.345,PhysRev.124.246,RevModPhys.64.649} which, in $(1+1)$-dimensions, is also known as the chiral Gross-Neveu ($\chi$GN) model~\cite{PhysRevD.10.3235}. This model involves $N$ flavours of Dirac fermions interacting via a $U(N)$ symmetric 4-Fermi coupling that can drive a spontaneous breakdown of chiral symmetry. 
Other paradigmatic low-dimensional models with either Abelian or non-Abelian continuous gauge groups are  Schwinger's QED$_2$~\cite{PhysRev.128.2425,COLEMAN1975267,COLEMAN1976239} and 't Hooft's QCD$_2$
~\cite{THOOFT1974461,WITTEN197957}, respectively. Both of them have been predicted to  host inhomogeneous chiral condensates at finite densities as well, e.g. see
~\cite{PhysRevD.19.1188,PhysRevD.50.1165,PhysRevLett.118.071601} and~\cite{PhysRevD.55.4920,Hayata2024}, respectively. 

Surprisingly, even if the $\chi$GN model is  a priori simpler than these low-dimensional gauge theories, 
consensus on the existence of  inhomogeneous chiral waves  has only been reached very recently. At zero chemical potential and vanishing temperatures,  large-$N$ methods predict a homogeneous complex-valued fermion condensate that breaks chiral symmetry, together with Goldstone bosons associated with phase fluctuations~\cite{PhysRevD.10.3235}. This large-$N$ prediction is in conflict with the Mermin-Wagner-Coleman  theorem~\cite{PhysRevLett.17.1133,Coleman1973}, and it can be shown that the correlation functions of this condensate actually decay with a power law of the distance that is inversely proportional to the flavour number $N$~\cite{WITTEN1978110,PhysRevLett.129.071603}. Hence, the $N\to \infty$ prediction is singular, as any finite number of flavours can only withstand quasi-long-range order even at $T=0$. This reconciles with the fact that the continuous chiral symmetry cannot be spontaneously broken for such low dimensions. At non-zero temperatures,  large-$N$ methods  again predict a finite region with a non-vanishing condensate and a restoration of the $U(N)$ symmetry at higher temperatures 
~\cite{PhysRevD.10.3956,PhysRevD.11.779}. This is again in conflict with the generic expectation that finite-$T$ long-range order cannot exist in (1+1)-dimensional systems~\cite{Peierls_1936}, which, in this case,  also applies to the discrete-symmetry  GN model~\cite{PhysRevD.12.2443}. 
For the continuum symmetry of the $\chi$GN model, very recent finite-$N$ results have shown that, indeed, the order only persists for lengths below a thermal wavelength proportional to $N$~\cite{Ciccone2024}, such that thermal long-range order is an artefact of the limit $N\to \infty$. At non-zero densities and temperatures,  initial large-$N$ studies only found 
non-vanishing homogeneous condensates~\cite{WOLFF1985303}. These predictions were later corrected by various  studies~\cite{PhysRevD.62.096002,PhysRevD.65.085040,PhysRevD.67.125015,PhysRevD.69.067703,PhysRevD.79.105012} showing that, in the large-$N$ limit, there are regions in the $(T,\mu)$  space in which it is favourable to form an inhomogeneous condensate,  which connects to the chiral spiral in QCD matter. In light of recent results based on non-Abelian bosonization~\cite{Ciccone2024}, these chiral spirals should also be suppressed at finite temperatures for lengths beyond a thermal wavelength, and for any finite flavour number away from the singular point $N\to\infty$.  We thus conclude that inhomogeneous chiral waves could only arise at $T=0$ in the $\chi$GN model, and only associated with quasi-long-range-order that vanishes at sufficiently long distances.

The prospect of finding long-range-ordered inhomogeneous condensates for the discrete GN model is more compelling, as the continuous symmetry is exchanged for a discrete one: the $\mathbb{Z}_2$ chiral symmetry~\cite{PhysRevD.10.3235}. At least at $T=0$, the spontaneous breaking of discrete chiral symmetry is not incompatible with general theorems~\cite{PhysRevLett.17.1133,Coleman1973}. 
In contrast to the $\chi$GN model, where predictions based on non-Abelian bosonization have led to the non-perturbative behaviour discussed above, the situation for discrete chiral symmetry is not yet settled.
First lattice studies also focused on  large-$N$~\cite{COHEN1983102,KARSCH1987289,Wenger_2006}, while more recent ones are addressing the discrete GN model with an even  $N$, where Monte Carlo does not encounter a sign problem~\cite{PhysRevD.101.094512,PhysRevD.102.114501}, {at least in principle}.

For fermions, the lattice discretisation of the $\mathbb{Z}_2$GN model opens a wide variety of possibilities that have been developed over the years to face the doubling problem~\cite{NIELSEN198120,NIELSEN1981173}. From the perspective of QFTs, finite-size scaling around a critical point or line of these lattice models should connect to the continuum QFT, giving predictions that are free of any lattice artefact or any remnant of the specific discretisation. This perspective changes when considering quantum simulators, as the lattice discretisation is physical, and one might be interested not only in critical points and continuum limits, but actually in the possible exotic phases of the specific lattice field theories. In fact, these QSs are real experiments with controllable and well-isolated many-body systems, and these phases are not artefacts but the experimental confirmation of non-trivial orders of matter. In this respect,  
Wilson's fermion discretisation~\cite{Wilson1977} is particularly interesting, as it  connects~\cite{KAPLAN1992342,GOLTERMAN1993219,PhysRevLett.108.181807} to symmetry-protected topological (SPT) insulators (see reviews~\cite{RevModPhys.82.3045,RevModPhys.83.1057,RevModPhys.88.035005}). 

In $(1+1)$ dimensions, Wilson fermions can be formulated on a cross-link graph, the so-called  Creutz ladder~\cite{PhysRevLett.83.2636,RevModPhys.73.119}, in which the spinor components get mapped onto the legs of a ladder that is pierced by an external flux. In the $\pi$-flux regime, one recovers a pair of Dirac fermions at low energies with different tunable Wilson masses, which can account for a non-zero topological invariant in the bulk and topological edge states in the boundaries. Adding density-density Hubbard interactions leads to a Wilson-type discretisation of the $\mathbb{Z}_2$GN model with a single flavour $N=1$~\cite{PhysRevX.7.031057,BERMUDEZ2018149,PhysRevB.99.125106,PhysRevB.106.045147}, and leads to an interesting interplay of SPT phases and homogeneous fermion condensates at zero temperature and density. 

\subsection*{Brief summary of the results}
In the direction of QCD matter at finite density and temperature, a recent study has argued that the $\mu=0$ SPT phase in the $N=1$ GNW model can survive a certain amount of thermal fluctuations~\cite{PhysRevLett.134.053002},\footnote{This study characterises how the soliton/kink excitations proliferate as temperature increases, such that the thermal topological order parameter vanishes beyond a critical temperature.  Using determinant quantum Monte Carlo with a single scalar auxiliary field, the authors calculate the flow of the critical lines separating the SPT phase from a trivial one as $T$ increases. In light of previous results at $T=0$~\cite{PhysRevX.7.031057,BERMUDEZ2018149,PhysRevB.99.125106,PhysRevB.106.045147}, it is expected that introducing a pseudoscalar auxiliary field would allow to complement the $T=0$ phase diagram with the parity-broken phase~\cite{BERMUDEZ2018149}, so-called Aoki phase in the context of lattice gauge theories~\cite{PhysRevD.30.2653}. At non-zero temperatures, however, this long-range order must be destabilised by thermal fluctuations ~\cite{Peierls_1936,PhysRevD.12.2443}, and one should find a thermal length above which the long-range condensates are distorted}. The goal of the present manuscript is to explore the other axis, and discuss the Gross-Neveu-Wilson (GNW) model at  $T=0$ but non-zero fermion densities, focusing on the single-flavour limit $N=1$ where quantum fluctuations can be more effective at destabilizing possible long-range-ordered phases. In particular, we explore the existence of chiral spirals and other possible inhomogeneous phases, and discuss the interplay of fermion condensation, solitons, and topology in this finite-density regime. Our results are derived by using a particular type of tensor network, the so-called matrix product states (MPS)~\cite{SCHOLLWOCK201196,RevModPhys.93.045003}. In addition to the aforementioned chiral spirals, which we identify for sufficiently strong interactions, we find that topological crystals also appear at weak interactions when doping the SPT vacuum with fermions or holes. We show that these crystals can be neatly understood in a dimerised limit of the GNW model, where the appearance of topological defects is associated to an extensive number of local conserved charges. These charges lead to a real-space manifestation~\cite{PhysRevLett.134.010411} of the phenomenon of Hilbert-space fragmentation~\cite{PhysRevX.10.011047,PhysRevB.101.174204,Moudgalya_2022},  partitioning the system in disconnected pieces separated by immobile defects that can host the extra doped fermions/holes due to a bulk-defect correspondence. The periodic distribution of these fermions/holes to minimise the overall energy leads to a topological crystal with a periodic charge arrangement.

For interactions beyond a certain critical value, we find a quantum phase transition to a non-topological parity-breaking phase characterised by a bulk pseudoscalar condensate. As we dope the system with an extra fermion/hole, we see that this condensate becomes inhomogeneous, and corresponds to a soliton/kink that interpolates between the two possible parity-broken values. Interestingly, due to the existence of the aforementioned conserved charges, we find that the kink can only have either positive or negative values of the topological charge, depending on the hole/fermion nature of the doping. As a consequence, as the fermion density is increased further, we find that the pseudoscalar condensate presents jumps at the fragmentation points, leading to a periodic distribution of equally-spaced single-charge anti-kinks. This contrasts with the standard situation in Yukawa-type fermion-boson QFTs~\cite{PhysRevD.13.3398}, in which one finds a periodic arrangement of kinks/anti-kinks with alternating +1/-1 topological charges. When abandoning the dimerised regime, the jumps in the pseudoscalar condensate are smoothened, and we also find that anti-kinks/kinks with the opposite charge start appearing at the interfaces.  Moreover, we find that the scalar condensate also starts to have non-vanishing values in these regions.  We show how, eventually, the finite-density ground state evolves into a periodic wave-like arrangement of the scalar and pseudoscalar condensates, which connects to  QCD's chiral spiral.

Before concluding this introduction, we want to emphasise that the phenomena discussed below could be tested in cold-atom QSs. In addition to the works~\cite{PhysRevX.7.031057,BERMUDEZ2018149,PhysRevB.99.125106,PhysRevB.106.045147}, there has been a recent interest in the digital Trotter-type quantum simulation of Gross-Neveu models~\cite{Czajka2022,PhysRevD.106.114515}. Following ~\cite{PhysRevX.7.031057,BERMUDEZ2018149,PhysRevB.99.125106,PhysRevB.106.045147}, we are interested instead in the prospects of analogue QSs based on cold atoms in Raman optical lattices~\cite{doi:10.1142/9789813272538_0001,PhysRevLett.121.150401,doi:10.1126/sciadv.aao4748,doi:10.1126/science.aaf6689,PhysRevResearch.5.L012006}. Tailoring the spin-dependent tunnelling and the atomic filling in these systems along the lines of~\cite{PhysRevResearch.5.L012006}, and also tuning the contact Hubbard-type interactions by Feshbach resonances,  it would be very interesting to explore experimentally the predictions based on the GNW model and the interplay of inhomogeneous condensates, topological phases, and strong correlations. 

This article is organised as follows: in Sec.\ref{sec:2} we introduce the GNW model, in its continuum and lattice versions, and review the phase diagram at zero density and temperature for the latter \cite{BERMUDEZ2018149}. Next, in Sec.\ref{Grand-canonical} we analyse the finite density regime along the symmetry line of the phase diagram, moving to the grand-canonical ensemble. We first use large-N methods, which assume homogeneous fermion condensates, and compare the results with those obtained through MPS-based algorithms, which do not rely on any theoretical assumptions and allow for assessing possible inhomogeneities. In light of the results shown, we move on to the canonical ensemble in Sec.\ref{sec:4}, fixing the total particle number and assessing the nature of the inhomogeneities observed, for both low and high densities. This is done by performing a local rotation on the original basis, addressing the problem on the `rung basis'. On this basis, we find an extensive number of conserved charges, which account for a strong Hilbert-space fragmentation. In this manner, we start by expanding the study developed for zero density to one particle/hole doping, using both conserved charges and MPS methods. We repeat the analysis for two-particle/hole doping, as well as for larger fillings, understanding the finite-density phenomenology by means of the Hilbert space fragmentation. At the end of the section, we move away from the symmetry line and show how the fermion condensates vary along the lattice for several points of the phase diagram, as a first probe for further inhomogeneous phases. We report our conclusions and outlook in Sec.\ref{sec:5}.

\begin{figure}
    \centering
    \includegraphics[width=0.75\linewidth]{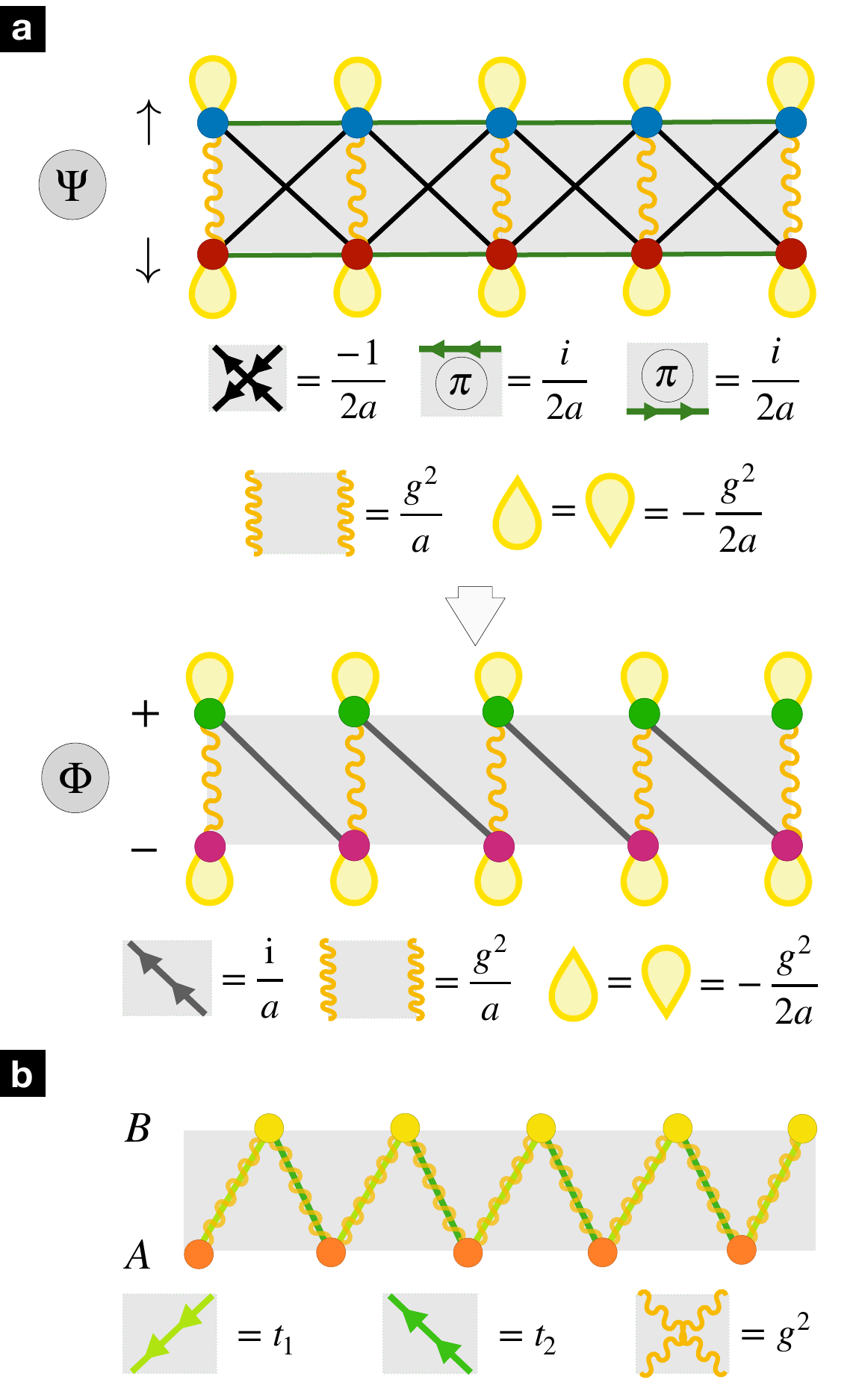}
    \caption{\textbf{Pictorial representation of the Gross-Neveu-Wilson (GNW) and Su–Schrieffer–Heeger (SSH) models.} \textbf{(a)} GNW model in the original $\Psi_{n,\sigma}$ (upper panel) and rung $\Phi_{n,s}$ (lower panel) bases, which are related via a local rotation, as explained in Sec.\ref{sec:4a}. By splitting the spinor indices $\sigma$ and $s$, the different terms contained in the Hamiltonians~\eqref{eq:GNW} and~\eqref{eq:rung_hamiltonian} can be depicted in 2-leg ladders, where the upper and lower legs correspond to the $\uparrow$ ($+$) and $\downarrow$ ($-$) indices of $\Psi_n$ ($\Phi_n$). The solid lines stand for the hopping terms, which in the $\Psi_n$ basis read $ \Psi^\dagger_{n, \sigma}\Psi_{n',\sigma'}+{\rm H.c.}$; the wavy ones correspond to the density-density interaction associated with the quartic term, $\sum_{\sigma} \Psi^\dagger_{n, \sigma} \Psi^\dagger_{n, \sigma'} \Psi_{n,\sigma'}\Psi_{n, \sigma}$; and the uniform single-orbital lobes are proportional to the number operators $\Psi^\dag_{n,\sigma}\Psi_{n,\sigma}$, with a global action identical to adding an interaction-dependent chemical potential $\mu(g^2)=g^2/2a$, relevant only when assuming a grand-canonical ensemble. \textbf{(b)} Connection to an interacting SSH-type model: the chain is divided into unit cells of two sites, usually named $A$ and $B$, so that interaction consists of the tunnellings and density-density repulsions between the nearest neighbour alternating species.}
    \label{fig:scheme_dimerised}
\end{figure}

\section{\bf The Gross-Neveu-Wilson (GNW)  model}\label{sec:2}

The Gross-Neveu model is a relativistic QFT that features 
$N$
flavours of  Dirac spinors $\Psi(x)=(\psi_1(x),\cdots,\psi_N(x))$, confined to a (1+1)-dimensional Minkowski spacetime~\cite{PhysRevD.10.3235}, with interactions governed by a four-fermion term invariant under a discrete $\mathbb{Z}_2$ chiral symmetry. In a Hamiltonian formulation, and up to an irrelevant constant in the half-filled case, 
\beq
\label{eq:GN_qft}
H=\int{\rm d}x\left(\overline{\Psi}(x)(-\ii\gamma^1\partial_x+m){\Psi}(x)-\frac{g^2}{2N}\big(\overline{\Psi}(x){\Psi}(x)\big)^2\right),
\eeq
where $g^2$ is the interaction strength, $m$ stands for the bare mass, and $\overline{\Psi}(x)=\Psi^{\dagger}(x)\gamma^0$ is the adjoint. Here, we have introduced the gamma matrices $\gamma^0=\sigma^z$ and $\gamma^1=\ii\sigma^y$, such that $\gamma^5=\sigma^x$ and the massless QFT would be invariant under the $\mathbb{Z}_2$ chiral transformation, corresponding to the axial rotation $\Psi(x)\mapsto\mathcal{R}_{\rm ax}\Psi(x)=\gamma^5\Psi(x)$. As discussed in~\cite{PhysRevD.10.3235}, this model shares key properties with higher-dimensional QCD, such as asymptotic freedom, dimensional transmutation, dynamical mass generation, and spontaneous chiral symmetry breaking. Regarding the latter, the model develops a non-zero fermion condensate for any non-zero interaction $g^2>0$, namely
\beq \label{eq:scalar_cond}
\sigma_0(x)=\langle\overline{\Psi}(x)\Psi(x)\rangle \,.
\eeq
In particular, this condensate is found to be invariant under translations $\sigma_0(x)=\sigma_0\,\,\forall x$, and invariant $\sigma_0\mapsto \sigma_0$ under parity transformations $\Psi(x)\mapsto\mathcal{P}\Psi(x)=\gamma^0\Psi(-x)$, i.e. a homogeneous scalar condensate $\sigma_0(g^2)$. This condensate is responsible for the spontaneous breakdown of the $\mathbb{Z}_2$ chiral symmetry for $m=0$, and endows the fermions with a non-perturbative dynamical mass that can be estimated in the large-$N$ limit~\cite{PhysRevD.10.3235}.

\begin{figure}
    \centering
    \includegraphics[width=0.9\linewidth]{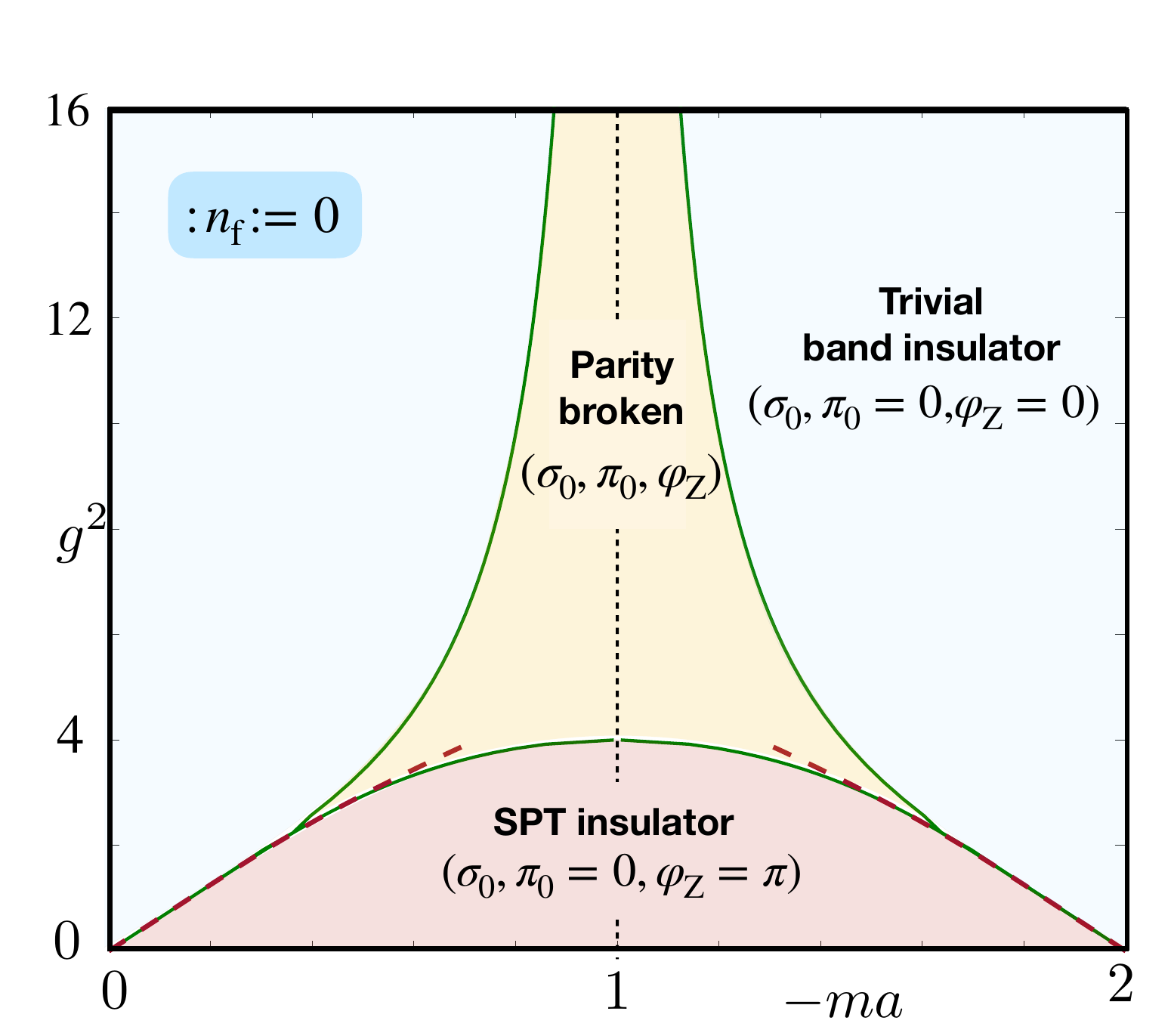}
    \caption{ \textbf{Schematic representation of the GNW phase diagram at zero temperature $T$ and density $:\!n_{\rm f}\!:$}: depending on the values of the bare mass $ma$ and the interaction strength $g^2$ there are three distinct phases: a SPT phase (shaded in red) characterised by the topological Zak phase $\varphi_{\rm Z}=\pi$ and a zero pseudoscalar condensate $\pi_0$; a parity broken (Aoki) phase (shaded in yellow) where $\pi_0$ is spontaneously broken and $\varphi_{\rm Z}$ adopts non-quantised values; and a trivial band insulator (shaded in blue) where the previous quantities are null. The scalar condensate $\sigma_0$ takes non-vanishing values in all the phase diagram except in the symmetry line $ma=-1$ (vertical dotted line). These three phases are delimited by second-order phase transitions (green lines). The red dashed lines indicate the SPT-Trivial critical lines, and follow from the condition $m(g^2)=0$ (left) and $m(g^2)=-2$ (right), with $m(g^2)=m+\sigma_0$.}
    \label{fig:scheme_phases}
\end{figure}

As argued in the introduction, we are interested in a Wilson-type lattice discretisation~\cite{Wilson1977,BERMUDEZ2018149} of this Hamiltonian field theory $x\to x_n=an$, $\Psi(x)\to\Psi_n$ {with $n\in\mathbb{Z}_{N_{\rm s}}$, $a$ the lattice spacing, and $L=aN_s$ the length of the lattice}. This lattice field theory reads
\beq
\label{eq:GNW}
H=a\sum_n\left(\left(\overline{\Psi}_n\mathbb{T}\Psi_{n+1}+{\rm H.c.}\right)+\overline{\Psi}_n\mathbb{M}\Psi_n-\frac{g^2}{2N}(\overline{\Psi}_n\Psi_n)^2\right),
\eeq
where $\mathbb{T}=(-{\ii}\gamma^1-r\mathbb{I}_2)/2a$, $\mathbb{M}=\left(ma+r\right)\mathbb{I}_2/a$, and $r$ is the dimensionless Wilson parameter, typically set to $r=1$. The  operators fulfill $\{\psi^{\phantom{\dagger}}_{n,\sigma,\alpha\phantom{\beta}\!\!\!},\psi^{\dagger}_{\ell,\tau,\beta}\}=\delta_{\alpha,\beta}\delta_{\sigma,\tau}\delta_{n,\ell}/a$, where $\alpha,\beta\in\mathbb{Z}_{N}$ ($\sigma,\tau\in\mathbb{Z}_2$) are flavour (spinor) indexes.

 As noted in the introduction, for $N=1$, one can show that the four-fermion terms reduces to a simple Hubbard interaction and, interpreting the spinor components as two legs in a ladder, the discretised lattice field theory can be connected to a cross-link Creutz-Hubbard ladder~\cite{PhysRevX.7.031057} (see upper panel of Fig.~\ref{fig:scheme_dimerised}{\bf (a)}). One expects to recover the physics of Eq.~\eqref{eq:GN_qft} in the vicinity of some critical point of the lattice model where the relevant length scale $\xi\gg a$. At zero temperature and density, the phase diagram of this model is depicted in Fig.~\ref{fig:scheme_phases}, which shows that the physics on the lattice is actually richer than the sole dynamical mass generation and the scalar condensate $\sigma_0$ predicted by the continuum QFT. First of all, except for the symmetry line $ma=-1$, one finds a non-zero scalar condensate $\sigma_0$ that renormalises the bare mass $m\to m(g^2)=m+\sigma_0$ leading to the portion of the critical lines highlighted as red dashed lines where {$m(g^2\searrow0)\to0$ or $m(g^2\searrow0)\to-2/a$}. These lines separate the red and blue regions, which host two distinct ground states/vacua, the difference of which can only be understood by considering their topological properties. The red area stands for an SPT insulator that not only has a non-zero scalar condensate, but also a quantised topological invariant $\varphi_{\rm Z}=\pi$ when additional symmetries are present. For $g^2=0$~\cite{BERMUDEZ2018149}, this invariant is the Zak phase~\cite{PhysRevLett.62.2747} that characterises the principal fibre bundle associated to the Bloch eigenstates within the Brillouin zone. For non-zero interactions $g^2>0$, one can connect the Zak phase to an interacting regime $\varphi_{\rm Z}(g^2)$ using the self-energy~\cite{PhysRevX.2.031008,PhysRevB.99.125106}, or otherwise resort to an operational definition related to the many-body position operator~\cite{PhysRevLett.80.1800} that can be extended to finite temperatures~\cite{PhysRevLett.134.053002}. In the blue area,  an inversion of the Wilson masses takes place,  making the topological invariant vanish $\varphi_{\rm Z}(g^2)=0$. The only remaining phase is the yellow one, which is separated from the rest by lines of second-order phase transitions. The ground state in this phase displays  a different fermion condensate 
\beq
\label{eq:pseudo_scalar_cond}
\pi_0(x)=\ii\langle\overline{\Psi}(x)\gamma^5\Psi(x)\rangle,
\eeq
which   is again found to be  invariant under translations $\pi_0(x)=\pi_0\,\,\forall x$, but this time odd $\pi_0\mapsto -\pi_0$ under parity. The homogeneous pseudoscalar condensate $\pi_0(g^2)$  thus signals the spontaneous breakdown of parity and, in this model, typically occurs in conjunction with a non-zero $\sigma_0(g^2)$ except along the symmetry line $ma=-1$. The yellow region is thus generally characterised by both a non-zero scalar and pseudoscalar fermion condensates, while the topological invariant is no longer quantised due to the symmetry breaking.

The goal of the following sections is to understand the physics of this model at finite fermion densities, which, for $N=1$, would result in a sign problem for Monte Carlo sampling. We will thus advocate for a large-$N$ grand-canonical approach and, eventually,  MPS numerical simulations both in the grand-canonical and canonical ensembles, in order to account for strong correlations and possible inhomogeneous phases. We will start by focusing on the symmetry line $ma=-1$, as it is at the same time the simplest and the more exotic regime regarding the possibility of finding inhomogeneous condensates and topological crystals, and connecting them to the phenomenon of Hilbert-space fragmentation. \vspace{0.6cm}

\section{\bf Grand-canonical ensemble and compressibility}\label{Grand-canonical}

\subsection{Large-\texorpdfstring{$N$}{Lg}  homogeneous condensates}
Let us start our study of the finite-density GNW model by inserting a non-zero chemical potential. 

\begin{figure*}
    \centering
    \includegraphics[width=1\linewidth]{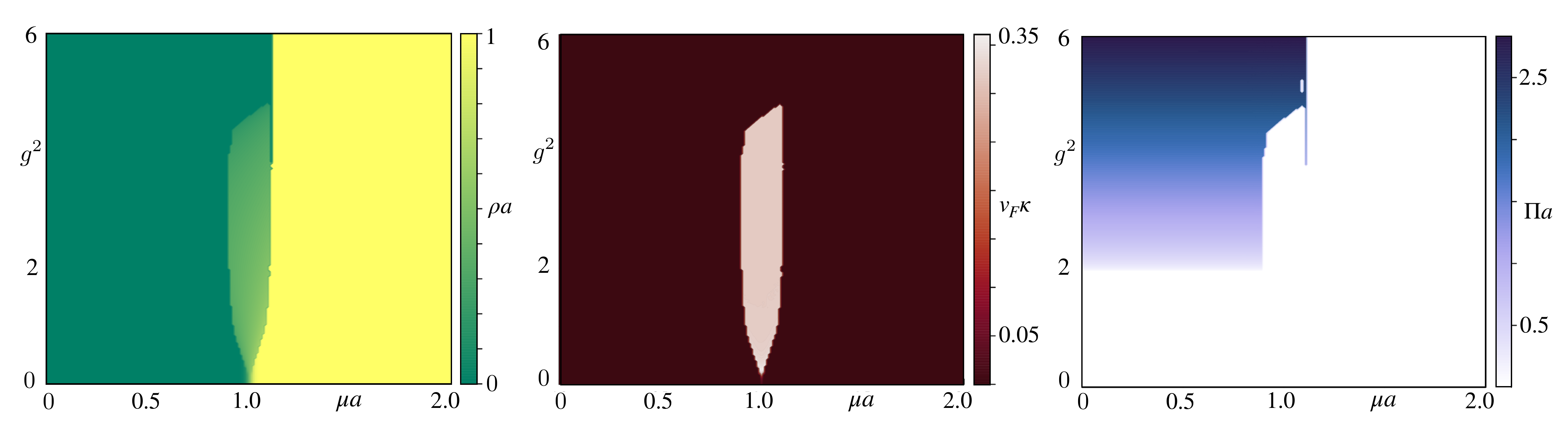}
    \caption{\textbf{Large-$N$ prediction for the GNW phase diagram near $\boldsymbol{ma=-1}$ .} Phase diagram in the $(\mu,g^2)$ plane obtained by minimising the large-$N$ effective potential on a 1024-site lattice assuming spatial homogeneity. The parameters are $\beta=100,\,ma=-0.999$.
    From left to right, it is shown the density $\rho$, the product $v_F\kappa$, and the pseudoscalar condensate $\Pi$. According to large-$N$ calculations, there is a bounded compressible phase that takes place at $\mu a=1$ in the free theory and widens for non-zero values of the interactions, ending at intermediate values of $g^2$. Regarding the pseudoscalar condensate, it takes non-vanishing values only inside the Aoki phase at zero density, a region that occurs at $g^2\geq 2$ according to the large-$N$ approximation.}
    \label{fig:condsmu_fig}
\end{figure*}
It is instructive first to set the scene using a large-$N$ approach assuming spatially-homogeneous condensation, and later on explore how this picture changes when one allows for inhomogeneous and strongly-correlated phases. To do this, we employ the effective potential $V_{\rm eff}(\Sigma,\Pi)$ approach outlined in~\cite{Ziegler:2021yua,Bermudez:2023nve,fulgadoclaudio2024interactingdiracfieldsexpanding},  which for $\mu=T=0$ was shown to improve upon the solution of the related gap equations~\cite{BERMUDEZ2018149}, allowing one to explore the large-$N$ version of the phase diagram depicted in Fig.~\ref{fig:scheme_phases}. We now  include  non-zero temperature and chemical potential, identifying the induced values of the scalar $\Sigma\sim\langle\bar\psi\psi\rangle$ and pseudoscalar $\Pi\sim \ii\langle\bar\psi\gamma_5\psi\rangle$ condensates
via  minimisation
\beq
(\sigma_0, \pi_0) = {\rm argmin} \{V_{\rm eff}(\Sigma, \Pi) :  \Sigma\in\mathbb{R}, \Pi \in\mathbb{R}\}
\eeq
Following the
derivation of the  extension of the effective potential   to $T,\mu\not=0$ \cite{Dashen:1974xz,Rosenstein:1988dj}, we find 
\begin{widetext}
\beq
\frac{V_{\rm eff}(\Sigma,\Pi)}{N}=\frac{1}{g^2}(\Sigma^2+\Pi^2)-\int_{
p}\Biggl(\max\big\{E_{\Sigma,\Pi}(p),\vert\mu\vert\big\}-\max\big\{E_{\boldsymbol{0}}( p),\vert\mu\vert\big\}
+T
\log\left(\frac{({1+\ee^{-\beta\vert E_{\Sigma,\Pi}+\mu\vert}})(1+\ee^{-\beta\vert E_{\Sigma,\Pi}-\mu\vert})}{{(1+\ee^{-\beta\vert
E_{\boldsymbol{0}}+\mu\vert})(1+\ee^{-\beta\vert
E_{\boldsymbol{0}}-\mu\vert})}}\right)
\Biggr),\label{eq:Veff}
\eeq
\end{widetext}
where $\mu$ is chemical potential, $\beta\equiv T^{-1}$ the inverse temperature,
and we consider 
a lattice with an even number of sites $N_{\rm s}$ such that the integral $\int_p$ over the first Brillouin zone is replaced by a sum over modes $p=2\pi n/aN_{\rm s}$ with $n=-N_{\rm s}/2+1,\ldots,N_{\rm s}/2$.   As discussed in~\cite{BERMUDEZ2018149}, the $\mathbb{Z}_2$-chiral-invariant 4-Fermi terms~\eqref{eq:GNW} must be decoupled into two possible condensation channels, such that the coupling strength appearing in this large-$N$ Hartree-Fock approximation $g^2\mapsto g^2/2$. 
The energy of the single-particle mode in the GNW model assuming spatially homogeneous $\Sigma,\Pi$ is given by 
\begin{equation}
\label{eq:bands}
    E_{\Sigma,\Pi}(p)=\sqrt{
    \frac{\cos^2(pa)}{a^2}+\left(\frac{1-\sin pa}{a}+m+\Sigma\right)^2+\Pi^2},
\end{equation}
with $E_{\boldsymbol{0}}(p)$ being the dispersion obtained with vanishing condensates $\Sigma=\Pi=0$. In the limit $T,\mu\to0$, equation~(\ref{eq:Veff}) recovers the result stated in Eq.~(C6) of \cite{fulgadoclaudio2024interactingdiracfieldsexpanding}~, after performing a Kawamoto-Smit rotation to connect the Creutz-Hubbard model to the standard Wilson discretisation, which takes $p\mapsto p-\frac{\pi}{2a}$, such that the Dirac points appear at $p_{\rm D}=\pm\pi/2a$. 

An estimate of the fermion density follows from 
\begin{equation}
\rho
=\int_{p }\Bigl(f\left(E_{\Sigma,\Pi}(p)\right)-\bar f\left(E_{\Sigma,\Pi}(p)\right)\Bigr),
\label{eq:density}
\end{equation}
where we used the Fermi-Dirac occupancy factors for particle/hole states
$f(E),\bar f(E)=1+\ee^{\beta(E\mp\mu)}$. Likewise,  the compressibility reads
\begin{equation}
\begin{split}
\kappa=\int_{p}\beta\Bigl(&f\left(E_{\Sigma,\Pi}(p)\right)(1-f\left(E_{\Sigma,\Pi}(p)\right))\\
&\bar f\left(E_{\Sigma,\Pi}(p)\right)(1-\bar
f\left(E_{\Sigma,\Pi}(p)\right))\Bigr).
\end{split}
\label{eq:chi}
\end{equation}

We numerically minimised the resulting effective potential~\eqref{eq:Veff} in search for the possibly non-zero homogeneous condensates $\sigma_0,\pi_0$, and Eqs.~(\ref{eq:density},\ref{eq:chi}) were subsequently evaluated 
on a system with $N_{\rm s}=1024$ and $\beta=100$, chosen to approximate the 
zero-temperature thermodynamic limit. We set the bare mass to $ma=-1+\epsilon$, with $\epsilon=10^{-3}$, close to but not exactly on the central symmetry line $ma=-1$ for $T=\mu=0$  (see Fig.~\ref{fig:scheme_phases}). This line is invariant under the symmetry $ma\to-2-ma$, $\Sigma \to-\Sigma$, provided that the scalar condensate vanishes, a symmetry that becomes manifest when integrating the energies~\eqref{eq:bands} over all momenta. The numerical minimisation yields the various quantities shown in Fig.\ref{fig:condsmu_fig}.

In the left panel of Fig.~\ref{fig:condsmu_fig}, we depict the fermion density $\rho$ as a function of $(\mu a,g^2)$, which shows that one remains in the half-filled state for a widegreen region in parameter space, corresponding either to the SPT or Aoki phases of the central line of Fig.~\ref{fig:scheme_phases}. In fact, looking at the pseudoscalar condensate $\pi_0$  of the right panel of Fig.~\ref{fig:condsmu_fig}, we find that the large-$N$ phase diagram shows a horizontal critical line that connects the critical point $g^2_c=2$ at $\mu=0$ already reported in~\cite{BERMUDEZ2018149}, to a new phase with a larger fermion density $\rho a\in(0,1)$ (pale green in the left panel). This finite-density feather-shaped region extends also towards a weakly-interacting regime, touching the axis at $\mu a=1$. At $g^2=0$, this chemical potential lies exactly at the energy of the flat bands $E_{\boldsymbol{0}}(p)=1/a$, such that higher chemical potentials correspond to a fully-filled ground state $\rho a=1$, where every site is occupied by a fermion, as shown in the Figure.  We note the transition as $g^2\to0$  is rapid but continuous --  the non-interacting system does not immediately saturate at $\mu a =1$.  In the central panel of Fig.~\ref{fig:condsmu_fig}{\bf (b)}, the pink shading of the feather-shaped  region denotes the product $v_{\rm F}\kappa$; multiplying the compressibility~\eqref{eq:chi} by the Fermi velocity $v_{\rm F}$ (estimated via the energy difference between modes straddling the Fermi energy $\mu$) corrects for the non-linearity of the GNW dispersion, so that the free-field result for continuum relativistic fermions $v_{\rm F}\kappa=1/\pi$ (which, e.g. with $v_{\rm F}=1$  follows from the free-field density
$\rho=\int_{-\mu}^\mu dp/2\pi$ for spinless fermions in one dimension) is recovered. These two results imply the existence of a Fermi surface with gapless excitations, i.e. the feather-shaped region is metallic in nature. In particular, note the factor $f(1-f)$ appearing in (\ref{eq:chi}) is only non-vanishing in the immediate vicinity of a Fermi surface. 

Although not shown in the plots, {we also find that} the scalar condensate $\sigma_0$ condensate is non-vanishing in the metallic phase, {growing with increasing $g^2$,} despite the proximity to the central line. The central line symmetry then implies that $\sigma_0$ must change sign as $ma$ decreases from $-1+\epsilon$ to $-1-\epsilon$, considering $\epsilon\to0^+$. Hence,  the effective potential predicts a first-order transition here, probably hinting at an instability of the homogeneous condensate phases.

As we shall see, allowing for spatial inhomogeneities in the condensates changes this phase diagram considerably. {There is a large body of works extending large-$N$ methods to allow for inhomogeneous condensates, going all the way from exactly-solvable cases~\cite{PhysRevD.69.067703}, to numerical approaches that postulate a specific form of the inhomogeneity~\cite{BUBALLA201539, narayanan2020phase}, e.g. plane wave,  in order to reduce the complexity of dealing with arbitrary auxiliary fields $\sigma(x),\pi(x)$ and the associated non-local effective action. As shown below, the inhomogeneities we find for the GNW model in the vicinity of the symmetry line are very different from such simple analytic modulations, and can even present sharp discontinuities due to a physical fragmentation of the Hilbert space. We thus note that generalising the large-$N$ approach for the particular problem at hand will not lead to very predictive inhomogeneous phases.

\begin{figure*}
    \centering
    \includegraphics[width=1\linewidth]{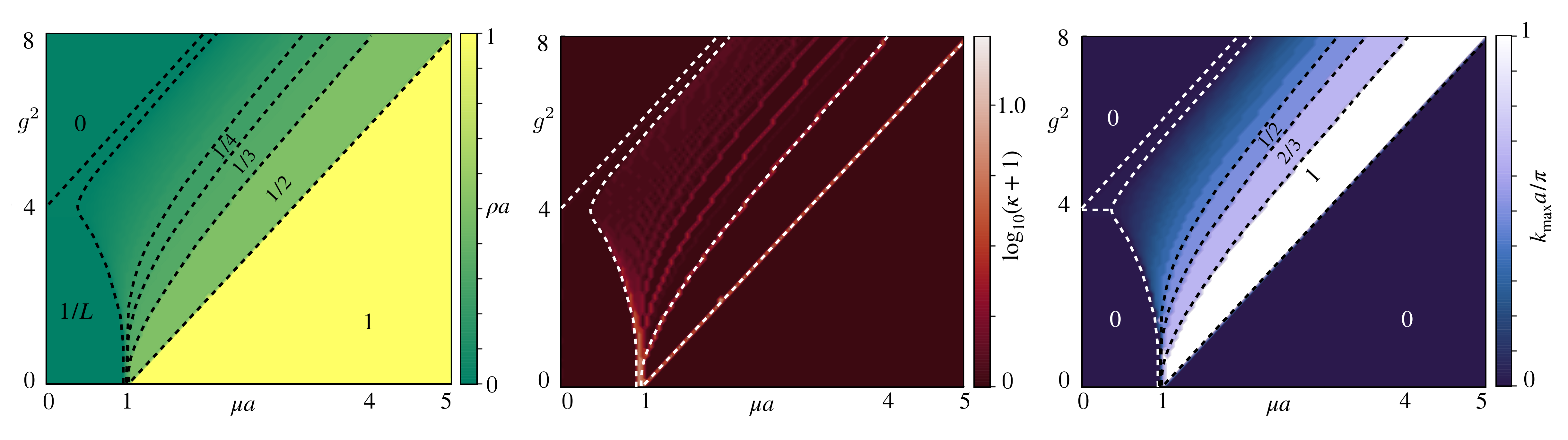}
    \caption{\textbf{Grand-canonical phase diagram of the GNW model at zero temperature}.  We have computed the ground state for several values of $(\mu a, g^2)$ along the symmetry line $ma=-1$ for chains composed of $N_s=128$ lattice sites. To characterise the phase diagram we have represented the average fermion density $\rho(\mu a, g^2)$ (left panel), the compressibility $\kappa(\mu a, g^2)$ (centre panel), and, in order to detect inhomogeneities in the pseudoscalar condensate, we work out its discrete Fourier transform (DFT) and retain the wavevector $k_{\rm max}(\mu a, g^2)$ with maximum amplitude (right panel). To pick out the relevant spatial modulations in $\pi_n$ from the ones associated with the boundary effects around $n=1$ and $n=N_s$, we have used a cutoff criterion: bearing in mind that $\pi_n$ is homogeneous in the bulk at zero density, we calculated the maximum amplitude of its DFT --extracted its mean-- at $g^2=4$, since the boundary effects are expected to be largest at the critical point. In this manner, only those $k_{\rm max}$ with greater amplitudes are considered. Clarified this, three qualitatively different regions are observed in the panels: first of all, the ones with either zero or unit density, which do not have spatial modulations on $\pi_n$. For intermediate values of $\rho$ there are two types of spatially-modulated phases as we increase the chemical potential: first, we find a compressible phase at low densities for $g^2>0$, while for larger ones plateaus around the fillings $\rho = (:\!n_{\rm f0}\!: \!a)^{-1}$ take place, being these more stable when corresponding to commensurate fillings. In all cases, we observe that the wavevector and the density are related by the expression $k_{\rm max}\approx 2\pi\rho$, which is exact for commensurate fillings while it may be approximate for the incommensurate ones.}
    \label{fig:mu_phases}
\end{figure*}

\subsection{Matrix-product-state  inhomogeneous phases}

Let us present in this subsection our approach based on matrix product states (MPS), which does not assume any specific inhomogeneity. At $T=0$, the thermodynamic properties of the system can be understood by finding  the ground state  of a modified grand-canonical Hamiltonian $H\to H_\mu=H-\mu N_{\rm f}$
\beq
\label{eq:e_0_mu}
\epsilon_0(\mu)= {\rm min}_{\ket{\psi}\in\mathcal{H}}\big\{\bra{\psi}\frac{1}{N_{\rm s}}(H-\mu N_{\rm f})\ket{\psi}\big\}
\eeq
where $H$ is the previous GNW Hamiltonian~\eqref{eq:GNW}, and $N_{\rm f}= a\sum_n\overline{\Psi}_n\gamma^0\Psi_n$ is the total fermion number operator. This energy can be understood as the corresponding $T\to 0$ limit of the grand-canonical potential 
$\Omega (\mu , T) = -T \rm{log} \mathcal{Z} (\mu , T)$, 
where 
$\mathcal{Z}(\mu,T)={\rm Tr}\{\ee^{-\beta H_\mu}\}$ is the partition function. 

We now use the MPS algorithm to approximate Eq.~\eqref{eq:e_0_mu} for different values of the chemical potential and interaction strength and, using finite differences, recover the average fermion density and compressibility
\beq
\rho(\mu,g^2)=\frac{1}{a}\frac{\partial\epsilon_0(\mu)}{\partial\mu}-\frac{1}{a},\hspace{2ex}\kappa(\mu,g^2)=\frac{\partial\rho(\mu)}{\partial\mu}.
\eeq
We note that we have subtracted the half-filled number of fermions of the Dirac sea from the fermion density $\rho$, and that the compressibility connects to the so-called quark number {susceptibility $\chi_q$} in the context of lattice QCD. The MPS approximation finds a variational upper bound to the energy~\eqref{eq:e_0_mu} by restricting the minimisation to a manifold $\ket{\Psi(\{\Gamma\})}\in\mathcal{M}_{\rm MPS}^{N_{\rm s},D}\subset\mathcal{H}$, where $\{\Gamma\}$ represents an array of $N_{\rm s}$ matrices of bond dimension $D$, contracted in the simplest tensor network: an MPS~\cite{SCHOLLWOCK201196,RevModPhys.93.045003}.

 \begin{figure*}
    \centering
    \includegraphics[width=0.75\linewidth]{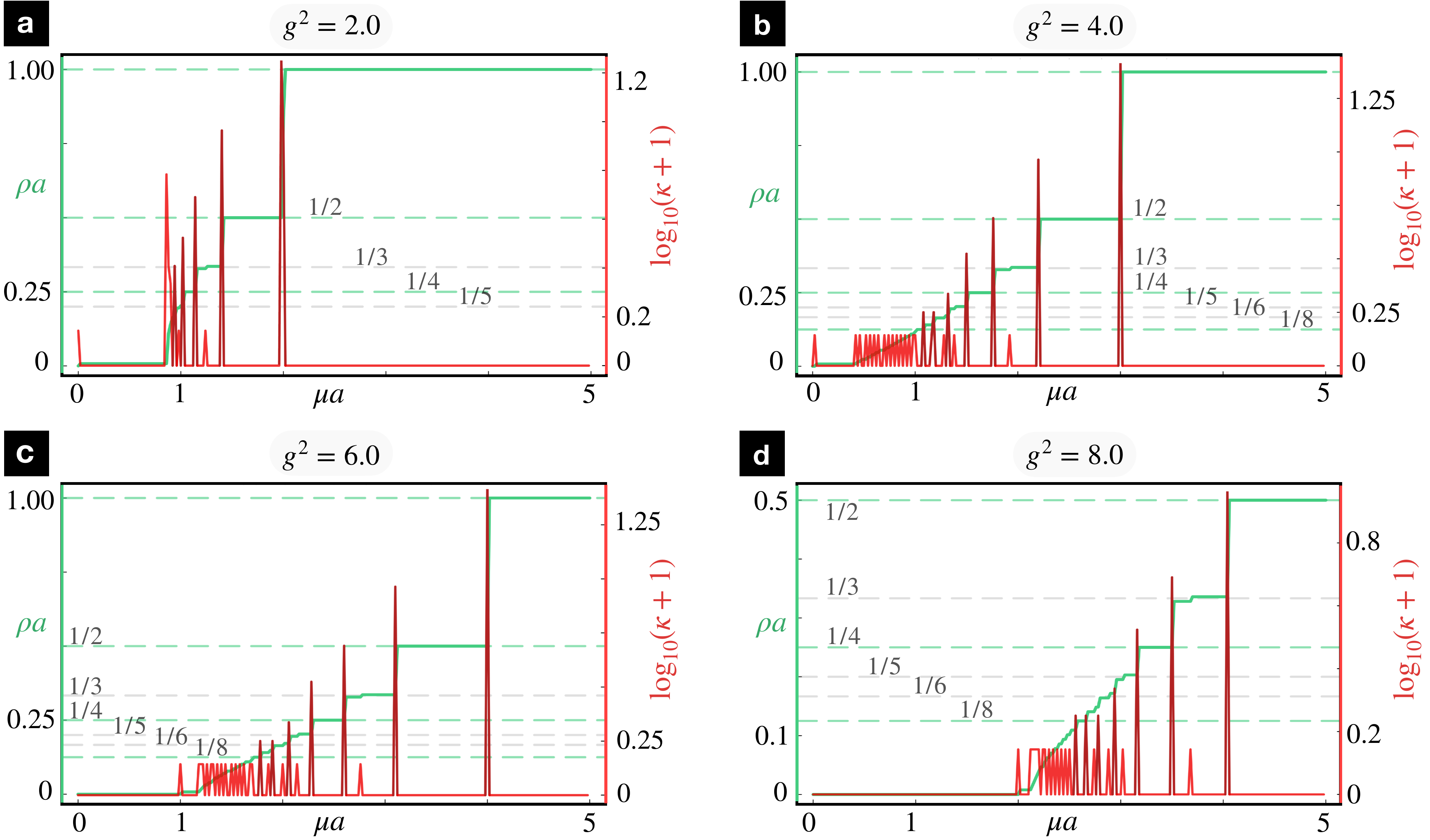}
    \caption{\textbf{Grand-canonical ensemble of the GNW model at zero temperature and $N_s=128$ lattice sites.} Density $\rho$ (green solid line) and compressibility $\kappa$ (red solid line) at \textbf{(a)} $g^2=2.0$, \textbf{(b)} $g^2=4.0$, \textbf{(c)} $g^2=6.0$ and \textbf{(d)} $g^2=8.0$. First of all, we observe that for $g^2\lesssim 4.0$ the density acquires a non-zero value at $\mu a \ll 1$, because of the existence of a zero-energy topological edge state which is populated as we turn on the chemical potential. Additionally, for $g^2>0$, once a given threshold in $\mu a$ is reached, the system enters a compressible phase, whose extension is $g^2$-dependent. This compressible phase is followed by an incompressible one consisting in plateaus around the densities $\rho = (:\!n_{f0}\!: \!a)^{-1}$ (dashed horizontal lines), being those associated with commensurable fillings (coloured in green) more stable.}
    \label{fig:rho_kappa_gsq}
\end{figure*}

Using a Jordan-Wigner transformation~\cite{Jordan:1928wi}, the fermionic Hilbert space can be expressed as $\mathcal{H}=\bigotimes_{n}\mathbb{C}^{2}$, such that a generic pure state is specified by $\mathcal{N}_{\rm p}=2\times(2^{2N_{\rm s}}-1)$ real parameters, up to an irrelevant global phase. In contrast, the MPS  manifold captures a low-entanglement corner of the Hilbert space using only $\mathcal{N}_{\rm MPS}=8(N_{\rm s}-2)D^2+16D-2(N_{\rm s}-1)D^2-2$ real parameters for open boundary conditions, which suffices to accurately approximate any ground state of a gapped local Hamiltonian because of the entanglement area laws~\cite{Hastings_2007, PhysRevB.73.094423,RevModPhys.82.277}. Thus, we will search the ground state variationally by means of the well-known density matrix renormalisation group (DMRG) algorithm~\cite{PhysRevLett.69.2863}, which optimises iteratively the energy taking entanglement as the relevant quantity to guide the process. {To perform these DMRG-based calculations we have used the  ITensor library \cite{ITensor, ITensor-r0.3}}. In what follows, we will work with MPS of $N_{\rm s}=128$ sites unless stated otherwise. Additionally, for the $ma=-1$ line, we will take $D_{\rm max}=50$, since in our tests the estimated compression error $||\ket{\psi}-\ket{\psi_{\rm trunc}}||$ committed in the state \cite{PhysRevB.73.094423} when reaching convergence during the optimisation process was of the order of $10^{-15}$, so one can assume that the physics is captured accurately.

In Fig.~\ref{fig:mu_phases}, we represent the possible finite-density phases obtained using our MPS-based algorithm. In the left panel, we represent the average fermion density $\rho$ as a function of the chemical potential and interaction strength, also known as the filling factor above half-filling. This contour plot shows in dark green (light yellow) the vacuum with zero density (saturated fermion density), and various other regions with commensurate and incommensurate fillings in between for $g^2>0$. In contrast to the corresponding $N\to\infty$ results of the left panel of Fig.~\ref{fig:condsmu_fig}, we appreciate that there is a much wider region of finite-density phases for a single fermion flavour $N=1$. All these phases emanate also from the $\mu a=1$ flat-band point as one increases both the interactions and changes the chemical potential. In contrast to the large-$N$ prediction of the saturated $\rho a=1$ region, we see that the critical chemical potential grows with $g^2/2$, which signals the energetic difference between the fully- and half-filled levels. In the large-$N$ limit, on the contrary, the kinetic energy scales with $N$ while the interaction energy is of order one, such that the energy will be dominated by degeneracy pressure. Once the bands are fully filled,  the  energy will no longer vary with $g^2$, such that the chemical saturation chemical  potential will not grow with the interactions, leading to a vertical line in accordance with Fig.~\ref{fig:condsmu_fig}.} 

Let us discuss further differences found for the $N=1$ case. At low chemical potential and $g^2<4$ we find a lobe at $\rho=1/L$, where $L=aN_{\rm s}$, in the lower left corner that connects to the SPT phase. Here,  the edge state that remained empty in the half-filled Dirac sea is now populated when considering a non-zero chemical potential. On the other hand, at $\rho=0$ and $g^2>4$ we find the parity-broken phase, where a pseudoscalar condensate~\eqref{eq:pseudo_scalar_cond} manifests a finite energy gap that inhibits the population of this extra fermion until the chemical potential exceeds the corresponding energy gap.
As $\mu a$ is increased, the system transitions to incompressible and compressible phases, which are readily identifiable by means of the compressibility displayed in the central panel. This quantity is zero in phases with a well-defined filling, but it diverges along the critical lines that separate two phases with different fillings. This allows us to identify the shaded black lines that surround the $\rho=1/L$ region, namely the straight dashed line that starts at $(\mu a, g^2)=(0,4)$, related to the already mentioned gap of the Aoki phase, and the curvy dashed line touching $(\mu a,g^2)=(1,0)$, point at which SPT phase with two filled edge states and the saturated density phase coexist. We see how, from this point, various lines emanate, surrounding phases with commensurate/incommensurate fillings.

To have a quantitative picture of these transitions, we represent in Fig.~\ref{fig:rho_kappa_gsq} various plots showing both density $\rho$ and compressibility $\kappa$  as one varies the chemical potential, considering  a fixed value of the interactions $g^2$. At low chemical potentials, e.g. Fig.~\ref{fig:rho_kappa_gsq}, we see an almost continuous succession of peaks in the susceptibility of variable extension, which shows that there is a region of gapless incommensurate phases connected by vanishingly small changes of the chemical potential. Conversely, at high chemical potentials, the density displays a sequence of wider and well-defined plateaus at fillings $\rho = (:\!n_{f0}\!: \!a)^{-1}$ --especially the commensurate ones, since otherwise there are transitions between the immediately below and above fractional fillings along the plateau--. These regions are delimited by sudden jumps of the compressibility that will diverge with system size.     Hence, we conclude that these fractional commensurate fillings correspond to gapped incompressible phases. 
{ A qualitatively  similar plot showing a staircase behaviour for $\rho(\mu)$ was obtained in numerical simulations of SU(2) QCD on a very small spatial lattice~\cite{Hands:2010vw}. In this case peaks in compressibility $\kappa$ (in this context known as quark number susceptibility $\chi_q$) coincide with colour-deconfined phases with a non-vanishing Polyakov loop}.

Finally, to shed more light on the nature of these partially-filled phases, we depict in the right panel of Fig.~\ref{fig:mu_phases} the momentum at which the discrete Fourier transform $f_k=(a/N)\sum_{n=1}^N e^{ikan}f_n$ of the pseudoscalar condensate~\eqref{eq:pseudo_scalar_cond} peaks. To avoid capturing the inhomogeneities coming from the boundaries, we have discarded all peaks with amplitudes lower than the one corresponding to the critical point $g^2=4$ at $\rho=0$, point at which the boundary effects are largest inside the well-known homogeneous regime. For non-zero and non-unity fillings with large system sizes $L\sim N_{\rm s}$, we find that the pseudoscalar condensate presents bulk inhomogeneities which, moreover, correspond to a periodic modulation with a well-defined Fourier peak at a non-zero momentum $k_{\rm max}$. For instance, as one lowers the chemical potential from the region with saturated fermion density $\rho=1/a$, the first inhomogeneous phase one encounters at $\rho=1/2a$ displays a peak at $k_{\rm max}=\pi/a$, signaling a dimerisation of the condensate which oscillates periodically with a two-site unit cell. This region is separated from the subsequent regions with $\rho=1/3a,1/4a$ fractional densities, which show a different modulation at $k_{\max}a=2\pi/3,2\pi/4$, each corresponding to a 3- and 4-site unit cell that repeats the condensate pattern periodically. In general, we find that $k_{\rm max}\approx 2\pi\rho$, a relation which is exactly(approximately) fulfilled at commensurate(incommensurate) fillings, a fact that will become clearer in the next sections.
In light of these results, we now move to the canonical ensemble in the following section, which will allow us to understand how the finite densities in the above compressible and incompressible phases distribute spatially.

\section{\bf Canonical ensemble and crystalline phases}\label{sec:4}

In the canonical ensemble, we work directly with the GNW Hamiltonian~\eqref{eq:GNW}, but focusing on a particular sector with a specific fermion number. $H$ is invariant under the global $U(1)$ transformation $\Psi_n\to\ee^{\ii\varphi}\Psi_n, \overline{\Psi}_n\to\ee^{-\ii\varphi}\overline{\Psi}_n$, such that the total fermion number $[H,N_{\rm f}]=0$ is a conserved quantity in this model, and the Hamiltonian can be decomposed in different blocks associated with all possible eigenvalues of $n_{\rm f}\in\sigma(N_{\rm f})=\mathbb{Z}_{2N_{\rm s}+1}$. Considering the fermionic Fock space $\mathcal{F}=\bigoplus_{n_{\rm f}\in\sigma(N_{\rm f})}\mathcal{F}_{n_{\rm f}}$, with $\mathcal{F}_{n_{\rm f}}={S}_-(\mathcal{H}^{\otimes {n_{\rm f}}})$, where $\mathcal{H}$ is the single-particle Hilbert space, and $S_-$ the anti-symmetrisation operator, the GNW Hamiltonian does not connect different subspaces $H:\mathcal{F}_{n_{\rm f}}\to \mathcal{F}_{n_{\rm f}}.$
An advantage of the MPS ansatz is that one can formulate the family of variational states restricted to one of these $\mathcal{F}_{n_{\rm f}}$ sectors in an exact manner \cite{singh2010tensor, singh2011tensor}. By choosing as the local basis, the eigenstates of the local fermion number operator, the symmetry is imposed by writing the network in terms of tensors that possess a block structure with respect to the $U(1)$ symmetry. This structure is imposed for higher-order tensors by assigning a ``charge" --in our case the fermion number-- for both physical and bond indices of the network, as well as a direction, usually depicted by an arrow, such that if the tensor has a total charge $Q$, its non-vanishing entries because of symmetry would be those satisfying the associated charge rules. For example, focusing on a tensor that preserves the charge/particle number, having then $Q=0$, the only non-vanishing entries by symmetry are those for which the ``charge" of the incoming indices is the same as the one of the outgoing indices. Bearing this in mind, we can construct an MPS, as well as the matrix product operator (MPO) representation of the Hamiltonian $H$, by means of these tensors, so that the total particle number is preserved during the optimisation process, assuming that the initial state has a well-defined total particle number. Concerning the possible values of the total fermion number, since the half-filled lattice corresponds to the vacuum of the Dirac field theory, we will define the fermion number as $:\!n_{\rm f}\!:\,= n_{\rm f} -N_{\rm s}$, where we recall that $N_{\rm s}$ is the number of lattice sites, such that the phase diagram in Fig.~\ref{fig:scheme_phases} corresponds to $:\!n_{\rm f}\!:\,=0$.

We will now start exploring other sectors with larger or smaller values of $:\!n_{\rm f}\!:$, which correspond to `doping' the system with fermions/holes with respect to the ground state, and particularise to the regime $ma=-1$, which is special for two reasons. First of all,  as discussed in~\cite{BERMUDEZ2018149}, the large-$N$ equations are invariant under the simultaneous transformation $ma\to -2-ma$, and $\sigma_0\to-\sigma_0$, such that the limit $ma=-1$ imposes a vanishing scalar condensate $\sigma_0=0$. In a Euclidean formulation, a similar condition leads to the so-called central-branch Wilson fermions, which have a translationally-invariant action that might be described at long wavelengths by two flavours of massless Dirac fermions even when $N=1$, allowing for a semi-positive definite Dirac determinant and a sign-error free formulation of the problem at $\mu=0$~\cite{10.1093/ptep/ptaa003}. One can show that there is an emerging symmetry along this central branch that, regardless of a large-$N$ approximation, forbids a non-zero scalar condensate, inducing an additive renormalisation of the bare mass~\cite{10.1093/ptep/ptaa003}. 

As we will see, this argument assumes a homogeneous condensate, and thus needs to be re-addressed in the presence of boundaries and finite fermion/hole densities  $:\!n_{\rm f}\!:\neq 0$. In the latter case, the semi-positivity of the Dirac determinant is no longer guaranteed, and one will most likely face a sign problem even when focusing on the central branch. In any case, we advocate for a MPS formulation, where the condition $ma=-1$ also has important consequences that become readily apparent in the Hamiltonian formulation.

\subsection{Rung basis, quasi-local conserved charges and Hilbert-space fragmentation}\label{sec:4a}

In this subsection, we show that the GNW model has an extensive number of conserved charges along the symmetry line $ma=-1$, which can be readily found by transforming the local field operators to a `rung basis' in the spinor components. We start by applying a Kawamoto-Smit phase rotation $\Psi_n\to\ee^{-\ii\pi n/2}\Psi_n, \,\,\overline{\Psi}_n\to\ee^{+\ii\pi n/2}\overline{\Psi}_n$~\cite{KAWAMOTO1981100} and, for simplicity, setting $N=1$ such that the flavour index disappears in the following. We  define the following rung operators
\beq
\label{eq:rung_basis}
\Phi_{n,+}=\frac{a}{\sqrt{2}}\Big(\Psi_{n,\uparrow}-\ii\Psi_{n,\downarrow}\Big),\,\,\,\Phi_{n,-}=\frac{a}{\sqrt{2}}\Big(\Psi_{n,\uparrow}+\ii\Psi_{n,\downarrow}\Big),
\eeq
which satisfy the canonical algebra $\{\Phi^{\phantom{\dagger}}_{n,s},\Phi^{\dagger}_{\ell,s'}\}=\delta_{s,s'}\delta_{n,\ell}$. The GNW Hamiltonian~\eqref{eq:GNW}  at $ma=-1$ adopts a  simpler form after the  phase rotation and dimer basis transformation
\begin{align}
\label{eq:rung_hamiltonian}
H &= \sum_{n}\left(\frac{\ii}{a} \Phi^{{\dagger}}_{n,+}\Phi^{\phantom{\dagger}}_{n+1,-}+{\rm H.c.}\right)+\frac{g^2}{a}\Phi^{{\dagger}}_{n,+}\Phi^{{\dagger}}_{n,-}\Phi^{\phantom{\dagger}}_{n,-}\Phi^{\phantom{\dagger}}_{n,+}-\nonumber\\ & -\frac{g^2}{2a} \left(\Phi^\dagger_{n,+}\Phi_{n,+} + \Phi^{\dagger}_{n,-}\Phi_{n,-}\right),
\end{align}
which has been depicted in Fig.~\ref{fig:scheme_dimerised} \textbf{(a)}. Here, the spinor components $\sigma\in\{\uparrow,\downarrow\}$ of the field are represented in a two-leg ladder by blue and red circles. The linear combination of operators in the rung according to Eq.~\eqref{eq:rung_basis} leads to the $s\in\{+,-\}$ operators represented by green and purple circles. Due to the nearest-neighbour tunnellings in Eq.~\eqref{eq:rung_hamiltonian}, the rung operators are only coupled in pairs, favouring the formation of dimers in the ground state. The Gross-Neveu quartic interactions, which are represented by a yellow wavy line, describe the repulsion of fermions residing on the same site but different $s$-orbital, which favours instead a vacuum where only the $s=+$ orbitals, or else the $s=-$ orbitals, are filled. This two-fold degeneracy is a result of a spontaneous breaking of the original parity symmetry, and accounts for the two possible values of the non-zero pseudoscalar condensate $\pi_0/|\pi_0|\in\{+1,-1\}$.  There is a competition between these two terms, leading to a quantum phase transition at $g^2=4$ at zero fermion density, which coincides with the intercept of the lower critical line with the symmetry axis in Fig.~\ref{fig:scheme_phases}.

The quadratic part of this model is reminiscent of the spinless Su-Schrieffer-Hegger (SSH) model~\cite{PhysRevLett.42.1698} (depicted in Fig.~\ref{fig:scheme_dimerised}\textbf{(b)}) in the limit of a static dimerised lattice. By flattening the zigzag structure, one can identify an odd-even tunnelling $t_1=0$ and an even-odd one $t_2=\ii/a$. Indeed, by moving away from this limit and exploring $ma\neq-1$, $(ma+1)/a\,\,\Phi^\dagger_{n,-}\Phi_{n,+}+{\rm H.c.}$ terms appear in the GNW Hamiltonian, so one can find a perfect analogy with the SSH model for $t_1=(ma+1)/a$ and $t_2=\ii/a$, which allows to neatly understand the presence of topological edge states for $-2<ma<0$ as the topological phase with $|t_1|<|t_2|$ (see the horizontal axis of Fig.~\ref{fig:scheme_phases}). In the limit $ma=-1$, one clearly sees from Eq.~\eqref{eq:rung_basis} that there are two unpaired orbitals at the edges $\hat{L}=\Phi_{1,-},\hat{R}=\Phi_{N_{\rm s},+}$, which thus represent the zero-energy states localised at the left and right edged in the SPT phase. 

This exact analogy cannot be carried on to the $g^2>0$ regime, as the spinless SSH model has been generalised by introducing density-density interactions between all nearest-neighbour pairs~\cite{PhysRevB.107.L201111,PhysRevB.110.165145}, and connects to a spinless Hubbard model~\cite{PhysRevB.12.3908} with dimerised tunnellings. In our case, nonetheless, the interactions couple only the odd rung orbital with the even one of the following dimer. Even if this can seem at first an unimportant lattice detail, we note that it can lead to completely different physics at long wavelengths. As an example, in the limit where $ma=0$ or $ma=-2$, where $|t_1|=|t_2|=1/a$, the dimerisation is absent and the interacting SSH model maps to an XXZ chain~\cite{PhysRevB.12.3908}, which is known to remain in a gapless critical phase for all nearest-neighbour interactions $g^2<|t_1|=|t_2|$, while an energy gap is opened via an infinite-order Kosterlitz-Thouless transition for interactions above $g^2>|t_1|$. This is very different from dynamical mass generation in our GNW model, which yields a second-order critical point for arbitrarily-small interactions $g^2\to 0^+$.

 This rung formulation will actually turn out to be the key to analyse the finite-density case as, in addition to the sectors associated with the above global $U(1)$ symmetry, it allows us to find an extensive number of quasi-local conserved charges consisting of the dimer number operators
 \beq
 \label{eq:dimer_number_ops}
 D^{\phantom{\dagger}}_{n}=\Phi^{{\dagger}}_{n,+}\Phi^{\phantom{\dagger}}_{n,+}+\Phi^{{\dagger}}_{n+1,-}\Phi^{\phantom{\dagger}}_{n+1,-},
 \eeq
 \beq
 [H, D_{n}]=0, \forall n\in\{1,2,\cdots, N_{\rm s}-1\}\,,
 \eeq
 whose spectra is $d_n\in\sigma(D_n)=\{0,1,2\}$ due to Pauli exclusion principle. In addition, at the edges, we can define two additional conserved charges $ [H, D_{0}]=[H, D_{N_{\rm s}}]=0$ related to the population of the edge states
\beq
\label{eq:edge_oparators}
D_{0}=\Phi^{{\dagger}}_{1,-}\Phi^{\phantom{\dagger}}_{1,-},\hspace{2ex} D_{N_{\rm s}}=\Phi^{{\dagger}}_{N_{\rm s},+}\Phi^{\phantom{\dagger}}_{N_{\rm s},+}.
\eeq
The spectrum of these operators is, in this case, $d_{\rm 0}\in\sigma(D_{\rm 0})=\{0,1\}$, $d_{N_{\rm s}}\in\sigma(D_{ N_{\rm s}})=\{0,1\}$.
Since all these operators commute with the Hamiltonian, we can decompose the corresponding Fock subspace $\mathcal{F}_{:n_{\rm f}:+N_{\rm s}}$ associated with a total number of fermions $:\!\!n_{\rm f}\!\!:$ into an extensive number of independent sectors $\mathcal{F}_{:n_{\rm f}:+N_{\rm s}}=\bigoplus_{\boldsymbol{d}}\mathcal{F}^{\boldsymbol{d}}_{:n_{\rm f}:+N_{\rm s}}$, where $\boldsymbol{d}=(d_{0},d_1,\cdots ,d_{N_{\rm s}-1},d_{N_{\rm s}})$ is subject to the 1-norm constraint on the total number of particles $\parallel\!\!\boldsymbol{d}\!\!\parallel_{\!1}=:\!n_{\rm f}\!:+N_{\rm s}$. We thus find a number of sectors increasing exponentially with system size for any value of the fermion tunnelling and Hubbard interaction. The emergence of these integrals of motion is thus connected to the phenomenon of strong Hilbert-space fragmentation ~\cite{PhysRevX.10.011047,PhysRevB.101.174204,Moudgalya_2022}. In fact, the above subspaces can be generated by the repeated action of the Hamiltonian~\eqref{eq:rung_hamiltonian} on a simple tensor product of Fock states with the total number of particles distributed among the dimers according to $\boldsymbol{d}$, and can thus be identified with Krylov subspaces~\cite{NANDY20251}. We note that the specific dimension of these subspaces can increase exponentially with $N_{\rm s}$ depending on the overall filling, e.g. for $:\!n_{\rm f}\!:=0$, we find ${ dim}(\mathcal{F}_{\rm gs})=2^{N_{\rm s}}$ as detailed below.

Let us also note that these quasi-local integrals of motion are related to an emergent subsystem symmetry~\cite{annurev_sym}, a transformation that is not strictly a local gauge symmetry but also not a global one, acting on fixed/rigid substructures. There is indeed a $U(1)$ transformation on a specific $n_0$ dimerised rung as $\Phi_{n_0,+}\mapsto\ee^{\ii\varphi}\Phi_{n_0,+}, \Phi_{n_0+1,-}\mapsto\ee^{\ii\varphi}\Phi_{n_0+1,-}$, and leaves the Hamiltonian invariant. Hence the symmetry line $ma=-1$ identified in the aforementioned large-$N$ studies is actually a very special regime with an extensive number of subsystem symmetries and a strong Hilbert-space fragmentation. In fact, the vanishing of the scalar condensate (\ref{eq:scalar_cond}) along the symmetry line predicted by large-$N$ methods can be understood by means of the subsystem symmetries: expanding the scalar condensate in the rung basis, $\sigma_n = \langle \Phi^\dagger_{n,+}\Phi_{n,-}-\Phi^\dagger_{n,-}\Phi_{n,+}\rangle/a^2$, we can see that it corresponds to the expectation value of an operator that not only breaks the quasi-local symmetry explicitly, but also maps any state contained in a given symmetry sector out to other sectors, as it exchanges particles between the adjacent dimers in a rung. Accordingly, for any eigenstate of the Hamiltonian $\sigma_n=0,\,\forall n$.

\textit{Conserved charges at zero density.} These conserved charges allow us to easily identify   the phenomenology happening at $:\!n_{\rm f}\!:=0$ along the symmetry line. First of all, we can pin down the subspace $\mathcal{F}_{\rm gs}$ that contains the ground state and other possible low-energy excitations, by inspecting the non-interacting limit, and then, assuming that there are no level crossings or first-order phase transitions,  conclude that the interacting ground state will lie in the same symmetry sector as $g^2$ is increased. This is consistent with our previous description ~\cite{BERMUDEZ2018149} of the phase diagram at zero fermion densities $:\!n_{\rm f}\!:=0$, displayed in Fig.~\ref{fig:scheme_phases}. For $g^2=0$, it is easy to see from Eq.~\eqref{eq:rung_hamiltonian} and Fig.~\ref{fig:scheme_dimerised} that the non-interacting half-filled ground state is two-fold degenerate, and can belong to any of the two sectors with $\boldsymbol{d}\in\mathbb{D}_{\rm gs}=\{(1,1,\cdots,1,0), (0,1,\cdots,1,1)\}$, such that $\mathcal{F}_{\rm gs}=\bigoplus_{\boldsymbol{d}\in \mathbb{D}_{\rm gs}}\mathcal{F}^{\boldsymbol{d}}_{N_{\rm s}}$ with $dim(\mathcal{F}_{\rm gs})=2\times 2^{N_{\rm s}-1}$. For $g^2<4$, each of these sectors contains one of the symmetry-protected zero-energy edge states populated, whereas for $g^2>4$ the two-fold degeneracy is caused by the spontaneous breakdown of parity $\pi_0/|\pi_0|\in\{+1,-1\}$. 

To understand the relation between each sector and these two possible values of $\pi_0/|\pi_0|$, let us introduce some useful constraints that follow from the conserved charges. Expanding the dimer~\eqref{eq:dimer_number_ops} and edge~\eqref{eq:edge_oparators} number operators in the original basis, we have that any eigenstate of the Hamiltonian fulfills the following constraints

\begin{figure*}
    \centering
    \includegraphics[width=1\linewidth]{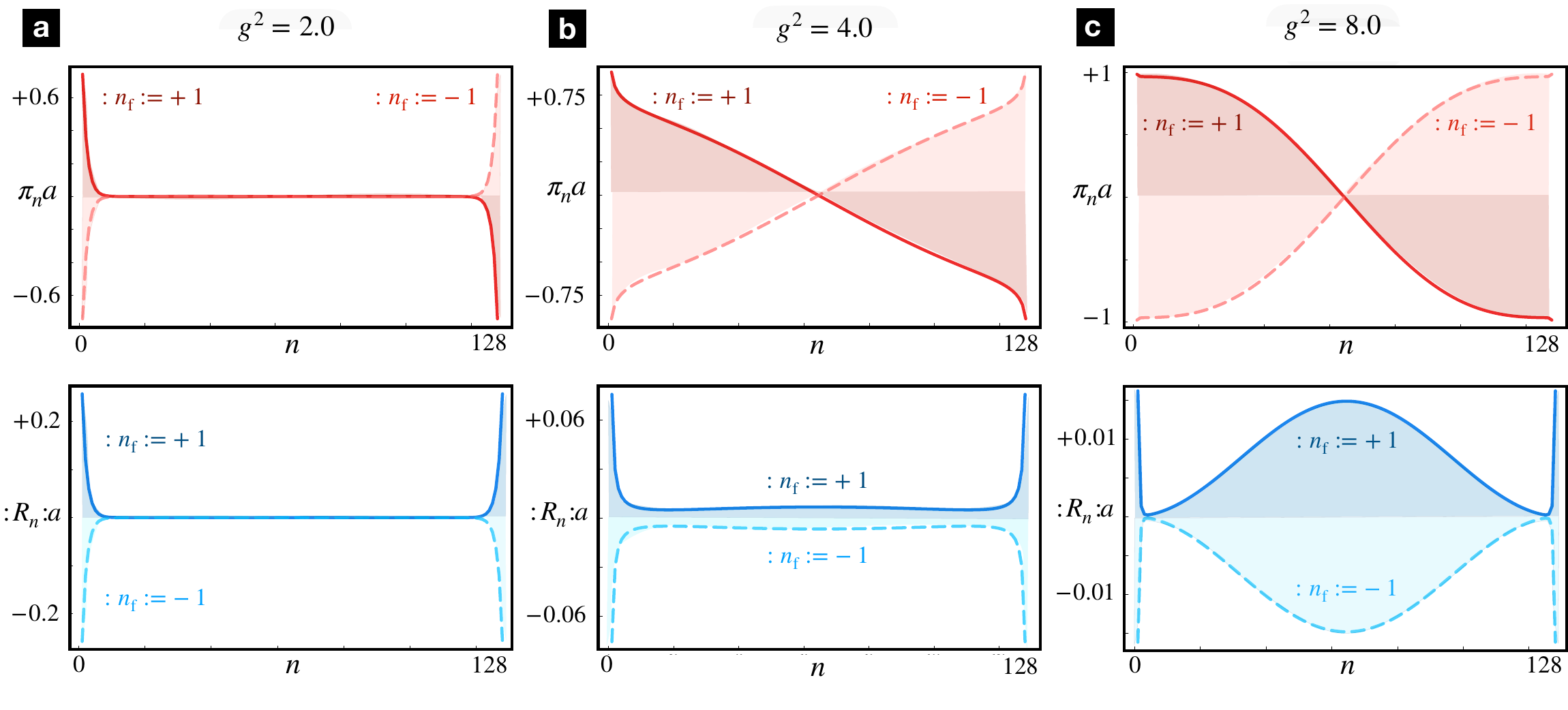}
    \caption{ \textbf{Inhomogeneities along the symmetry line $\boldsymbol{ma=-1}$ at $\mathbf{:\!n_{\rm f}\!:=+1}$.} Pseudoscalar condensate $\pi_n$ (upper panels) and rung density $:\!R_n\!:$ (lower panels) for one fermion/hole doping at \textbf{(a)} $g^2=2.0$, \textbf{(b)} $g^2=4.0$ and \textbf{(c)} $g^2=8.0$. Focusing on the fermion doping, we can observe that, inside the SPT phase $\pi_n$ is only non-vanishing around the edges, where the charge is concentrated due to the population of the exponentially-localised topological edge states. On the contrary, in the Aoki phase the pseudoscalar condensate shows an anti-kink profile, interpolating between the two possible values at the vacuum, $\pm |\pi_0|$, being the charge accumulated at the centre of the anti-kink. These two clearly different behaviours are interpolated around $g^2=4.0$, which corresponds to the critical point at zero density. The hole doping scenario is recovered by changing excess charge to deficit and inverting chirality.}
    \label{fig:pi_n+1}
\end{figure*}

\beq
\label{eq:conserved_charges}
\begin{split}
d_0&=\frac{a}{2}\left(R_1+ \pi_1\right),\\
d_n&=\frac{a}{2}\left( R_n+ R_{n+1}- \pi_n+\pi_{n+1}\right),\forall n\in\{1,\cdots ,N_{\rm s}-1\},\\
d_{N_{\rm s}}&=\frac{a}{2}\left(R_{N_{\rm s}}- \pi_{N_{\rm s}}\right),
\end{split}
\eeq
where we have introduced the total fermion density on pairs neighbouring rungs, and  discretised the derivative of the pseudoscalar condensate, which depend on
\beq
\begin{split}
R_n&=\langle \Psi^{{\dagger}}_{n,\uparrow}\Psi^{\phantom{\dagger}}_{n,\uparrow}\rangle+\langle\Psi^{{\dagger}}_{n,\downarrow}\Psi^{\phantom{\dagger}}_{n,\downarrow}\rangle, \\
\pi_n&=\ii\langle \Psi^{{\dagger}}_{n,\uparrow}\Psi^{\phantom{\dagger}}_{n,\downarrow}\rangle-\ii\langle\Psi^{{\dagger}}_{n,\downarrow}\Psi^{\phantom{\dagger}}_{n,\uparrow}\rangle.
\end{split}
\eeq

These relations allow us to iteratively obtain  two possible expressions for the pseudoscalar condensate
\beq
\label{eq:conserved_charges_pi_condensate}
\begin{split}
\pi_{n} &= \frac{2}{a}\sum_{k=0}^{n-1} d_k - 2\sum_{k=1}^{n-1} R_k - R_{n},\\
\pi_{n} &= -\frac{2}{a}\sum_{k=n}^{N_s} d_k + 2\sum_{k=n+1}^{N_s} R_k + R_{n}.
\end{split}
\eeq
At zero fermion density  $:\!n_{\rm f}\!:=0$, we have previously found that the particle density in the bulk is translationally invariant, so that  $R_n=R_{n+1}=1/a$~\cite{BERMUDEZ2018149}. Since $d_n=1$ in the bulk, we can conclude from the equations above that $\pi_n=\pi_{n+1}$, such that there are no bulk inhomogeneities in the fermion condensate, which is consistent with the typical large-$N$ assumptions. 

Let us now focus on the effect of the edges, where one expects possible boundary inhomogeneities. 
These  effects are captured by the sums on $R_k$ in the form of a charge excess $q$ over the homogeneous density for the left-most  boundary. Carrying these sums to the $n$-th bulk site, namely $\sum_{k=1}^{n-1} R_k = (n-1 + q)/a$ and $\sum_{k=n+1}^{N_s} R_k = (N_s - n - q)/a$, we get
\beq
\begin{split}
\pi_{n} &= \frac{2}{a}  \left(d_0 - q -\frac{1}{2}\right) = \frac{2}{a}  \left(  q + \frac{1}{2}-d_{N_s}\right),
\end{split}
\label{eq:bulk_pis_boundary}
\eeq
such that the bulk condensate attains a value of  $\pi_n = (-2q + 1)/a =: \pi_0$ for $(d_0,d_{N_{\rm s}})=(1,0)$, and $\pi_n = (2q - 1)/a = -\pi_0$ when fixing $(d_0, d_{N_{\rm s}})=(0,1)$. 
In the free case $g^2=0$,  each dimer contains a completely-localised particle. Since the quadratic terms in Eq.~\eqref{eq:rung_hamiltonian} do not favour any orbital within the dimer, we expect each to contribute with $1/2a$ to the density. In this manner, the charge excess associated with the  
edge states, 
$n\in\{1,N_s\}$, is $q=1/2 + 1 - 1=1/2$ (left edge state populated) or  $q=-1/2$ (right edge state populated). Substituting in Eq.~\eqref{eq:bulk_pis_boundary}, we find that $\pi_n=\pi_0=0, \; \forall n\in\{2,\cdots, N_s-1\}$, such that there is no bulk pseudoscalar condensate in this regime. On the other hand, for $g^2\gg1$, the strong repulsion between particles on a given rung enforces to populate only one of the orbitals $\Phi_{n,s}$, leaving the other one empty. We can thus ascertain that there will be a single fermion per rung $R_n=1/a\;\forall n\in\{1,\cdots, N_s\}$, such that $q=0$ and $\pi_n=+1/a$ for the sector $(d_0, d_{N_s})=(1,0)$, while $\pi_n=-1/a$ for $(d_0,d_{N_{\rm s}})=(0,1)$. Bearing in mind these limits, the qualitative behaviour along the symmetry line is as follows: as we turn on the interactions, the repulsion between orbitals of the same rung induces a delocalisation of the edge states from the boundaries. For small interactions, this delocalisation has a finite support, so that the excess charge $q$ is still $1/2$, and $\pi_n=0$ in the bulk. At $g^2=4$ the tails of the edge states reach the middle of the chain so that the charge excess of one of the boundaries partially compensates the lack of charge of the other one and $q$ starts to decrease, giving rise to a non-zero $\pi_n$ in the bulk for $g^2>4$. Finally, if we continue increasing the interactions the repulsion penalises the boundary effects, and the system tends to be completely homogeneous, such that $q$ tends to zero. This intuitive picture of the symmetry line, together with the Eqs.~\eqref{eq:conserved_charges} and \eqref{eq:conserved_charges_pi_condensate}, will be very useful when discussing the fermion/hole doping in the next subsections.

\subsection{Inhomogeneous condensates: doping the GNW model with one, two, and many fermions.}

\subsubsection{ Ground state topological phases for an extra fermion above half filling}

Let us now discuss how these constraints allow to predict possible inhomogeneities of the condensate upon doping $:\!n_{\rm f}\!:\neq0$. We start by considering  $:\!n_{\rm f}\!:=\pm 1$, namely doping the system with a single fermion/hole. The sector that contains the ground state can again be easily identified by looking at the non-interacting limit, and is either $\boldsymbol{d}=(1,1,\cdots,1,1)$ for the extra fermion, or $\boldsymbol{d}=(0,1,\cdots,1,0)$ for the extra hole. In this case, the previous two-fold topological or symmetry-breaking degeneracies are lost upon doping.

We start by outlining some generic differences in the relevant quantities compared to the zero-density scenario. First, since both edge states have the same occupation number, $d_0=d_{N_{\rm s}}$, and $d_n=1$ in the bulk, the rung density is an even function with respect to the centre of the chain, namely $R_n \to R_{N_{\rm s}+1-n}$. Because of this, according to Eq.~\eqref{eq:conserved_charges_pi_condensate} the pseudoscalar condensate is an odd-function $\pi_n \to -\pi_{N_{\rm s}+1-n}$, so parity symmetry is restored for these two sectors. Additionally, as will become clearer throughout this section, for $g^2\neq 0$ the repulsion between orbitals of the same rung delocalises the extra fermion/hole in either the boundaries or the bulk, so that for $:\!n_{\rm f}\!:=+1$ ($:\!n_{\rm f}\!:=-1$) the rung occupancies will fulfill $R_n+R_{n+1}\geq 2/a$ ($R_n+R_{n+1}\leq 2/a$), $\forall n$. Using Eq.~\eqref{eq:conserved_charges} and $d_n=1$ in the bulk, this implies that the fermion condensate becomes a monotonically decreasing/increasing function when doping the system with a fermion/hole
\beq
\label{eq:monotonous}
\begin{split}
:\!n_{\rm f}\!:=+1,\hspace{2ex} \pi_n>\pi_{n+1},\forall n,\\
:\!n_{\rm f}\!:=-1,\hspace{2ex} \pi_n<\pi_{n+1},\forall n.
\end{split}
\eeq

With this clarified, we will discuss in the following the inhomogeneities that appear along the symmetry line from small to large interactions, by inspecting both the pseudoscalar condensate and rung density, the latter redefined to capture the difference with respect to the half-filled bulk behaviour, $:\!R_n\!: \equiv R_n -1/a$. These inhomogeneities will be tested through our MPS algorithms, mainly for the three representative interactions $g^2\in\{2,\,4,\,8\}$, values associated with the SPT phase, critical point, and parity broken phase at zero density, respectively. Following the intuition developed for $:\!n_{\rm f}\!:=0$, we have that, for small interactions, the two edge states are exponentially localised at the boundaries, being both populated or emptied for $n_{\rm f} = +1$ and $n_{\rm f}=-1$, respectively. In both cases, if we repeat the same reasoning as for zero density, we would get that $\pi_n=0$ in the bulk, while it does not cancel near the boundaries due to the associated charge excess $q$, whose spatial distribution is captured by $:\!\!R_n\!\!:$. This qualitative picture is aligned with our results at $g^2=2$, shown in Fig.~\ref{fig:pi_n+1} \textbf{(a)}, where we observe that both $\pi_n$ (upper panel) and $:\!\!R_n\!\!:$ (lower panel) cancel in the bulk, while due to the occupation number of the edge states, $:\!\!R_n\!\!:$ adopts positive (negative) values near the boundaries for $:\!\! n_{\rm f}\!\!:=+1$ ($:\!\! n_{\rm f}\!\!:=-1$), with the consequent decreasing (increasing) behaviour for $\pi_n$.

For $g^2=4$, according to our knowledge for $\rho=0$, the delocalisation of the edge states reaches the middle of the chain, so that for $g^2\geq 4$ a non-trivial bulk behaviour takes place. In order to better understand this intermediate regime, let us focus on the $g^2\gg1$ situation first. At these coupling values, we are deep inside the zero-density parity-broken phase, where the pseudoscalar condensate takes two possible values in the bulk, $\pm |\pi_0|$. As shown in Fig.~\ref{fig:pi_n+1}\textbf{(c)} for $g^2=8$, we find that, when doping the system with a fermion, the ground state interpolates from $\pi_1/|\pi_0|=+1$ to $\pi_{N_{\rm s}}|\pi_0| =-1$ via a continuous decreasing function as one moves from left to right, leading to a kink-type solution which vanishes exactly at the middle of the chain $n_0=(N_{\rm s} +1)/2$. This profile is reminiscent of a soliton-type solution, and the fact that it vanishes at the centre of the chain is responsible for restoring the parity symmetry that was spontaneously broken at half filling. When doping with a hole, the description is analogous, but the accumulation of charge turns into depletion, and the chirality is reversed.

In the context of  semi-classical solitons  in non-linear field theories with spontaneous symmetry breaking~\cite{Rajaraman:1982is}, one can assign a single-valued topological charge to such solitons, which in our case can be expressed as 
\beq
Q^{:n_{\rm f}:=+1}_\pi=\frac{1}{2|\pi_0|}\int{\rm d}x\partial_x\pi(x)=\frac{1}{2|\pi_0|}(\pi_{N_{\rm s}}-\pi_1)=-1.
\eeq
In contrast to the standard situation in QFTs, where  $\mathbb{Z}_2$ solitons can display positive and negative charges,  typically refereed to as kinks or anti-kinks, we find that only anti-kinks are stabilised in the ground state of the GNW model when doping with an extra fermion. This is a consequence of the monotonously decreasing character of the fermion condensate~\eqref{eq:monotonous} that was derived above using the conserved charges. Since the condensate cannot increase for this particular filling and charge sector, we cannot find a succession of kinks and anti-kinks with opposite charges. As one observes in Fig.~\ref{fig:pi_n+1} \textbf{(c)}, kinks can only be found when doping with a hole. One obtains  the predicted monotonously increasing condensate, leading in this case to a positive topological charge
\beq
Q^{:n_{\rm f}:=-1}_\pi=\frac{1}{2|\pi_0|}\int{\rm d}x\partial_x\pi(x)=\frac{1}{2|\pi_0|}(\pi_{N_{\rm s}}-\pi_1)=+1.
\eeq

\begin{figure}
    \centering
    \includegraphics[width=0.8\linewidth]{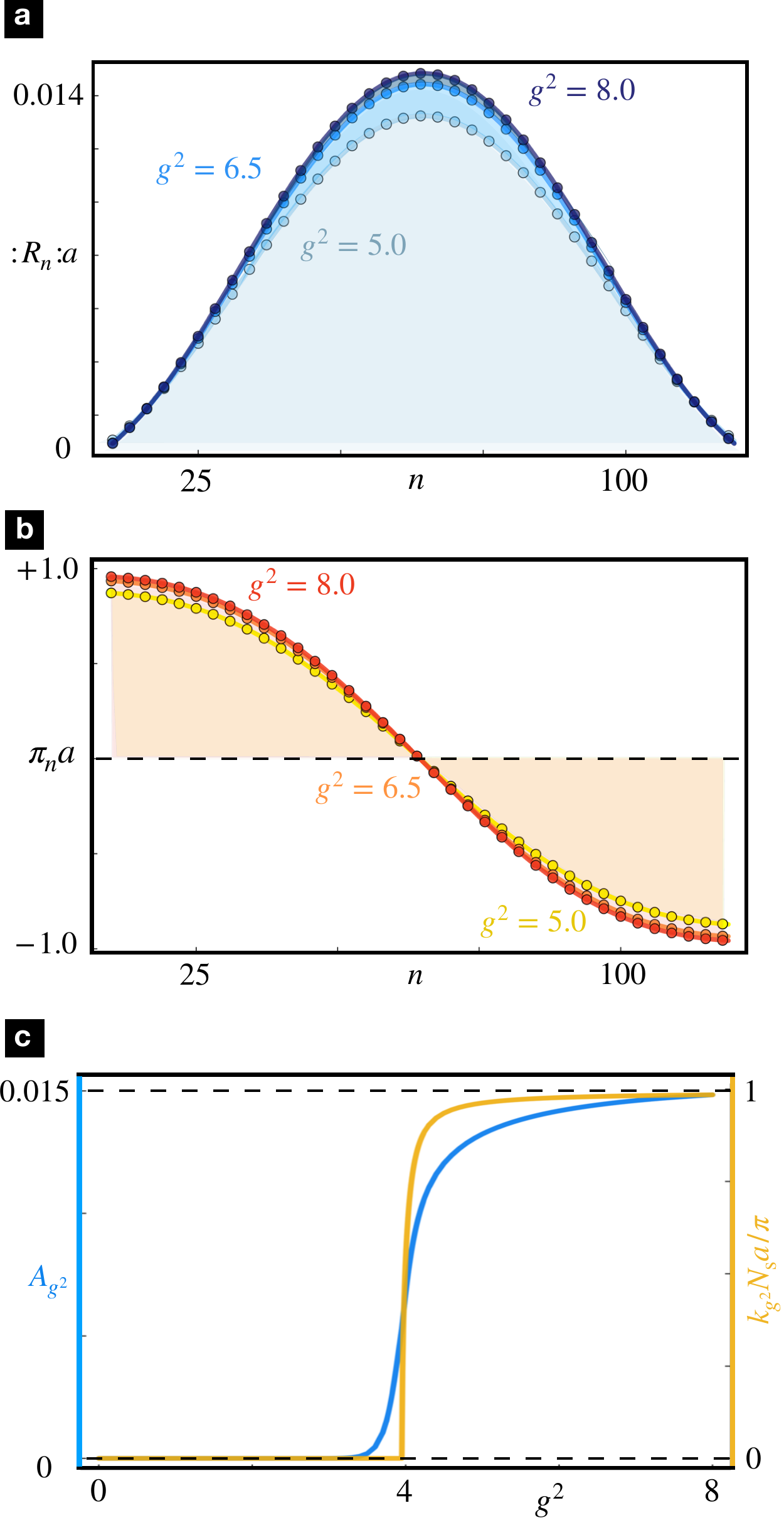}
    \caption{\textbf{Anti-kink fits at $\mathbf{:\!n_{\rm f}\!:=+1}$.} \textbf{(a)} Fit of the rung density $:\!R_n\!:$ data (points) to Eq.~\eqref{eq:Rn_fit} (lines) at $g^2=8.0$ (dark blue), $g^2=6.5$ (blue) and $g^2=5.0$ (light blue) for the ground state in the bulk, being the fitted parameters $(A_{8.0}, \;k_{8.0}N_sa/\pi)=(1.484\cdot 10^{-2},\,0.984)$, $(A_{6.5},\; k_{6.5})=(1.442\cdot 10^{-2},\, 0.980)$ and $(A_{5.0},\; k_{5.0})=(1.323\cdot 10^{-2},\, 0.962)$. \textbf{(b)} Fit of the pseudoscalar condensate $\pi_n$ data (points) to Eq.~\eqref{eq:pi_fit} (lines) at $g^2=8.0$ (red), $g^2=6.5$ (orange) and $g^2=5.0$ (yellow) for the ground state in the bulk, being the fitted parameters the same as for $:\!R_n\!:$. \textbf{(c)} Fitted parameters $A_{g^2}$ (blue line) and $k_{g^2}$ (yellow line) as functions of $g^2$: from strong to weak interactions the wavevector $k_{g^2}$ decays rapidly from the saturated value $\pi/N_s a$ to zero around the point $g^2=4.0$, what suggests a phase transition at that point also for the non-zero density regimes. Conversely, the amplitude decays more slowly.}
    \label{fig:rung_density_fits}
\end{figure}

\begin{figure}
    \centering
    \includegraphics[width=1.0\linewidth]{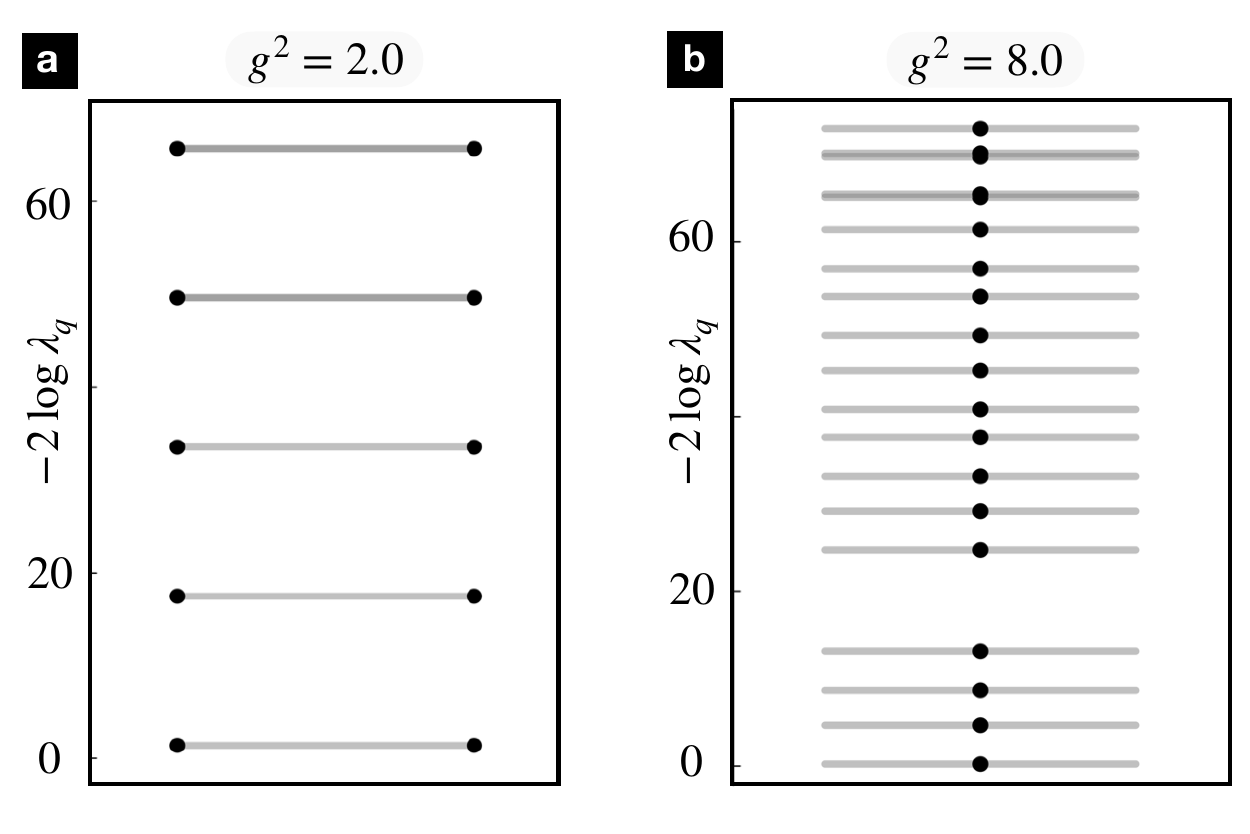}
    \caption{ \textbf{Entanglement spectrum for a single doped fermion.} \textbf{(a)} At $g^2=2.0$ it shows a double-degeneracy, as expected for a SPT phase, which is totally lost at $g^2=8.0$, \textbf{(b)}, deep inside the solitonic phase.}
    \label{fig:ES_N+1}
\end{figure}

\begin{figure*}
    \centering
    \includegraphics[width=0.75\linewidth]{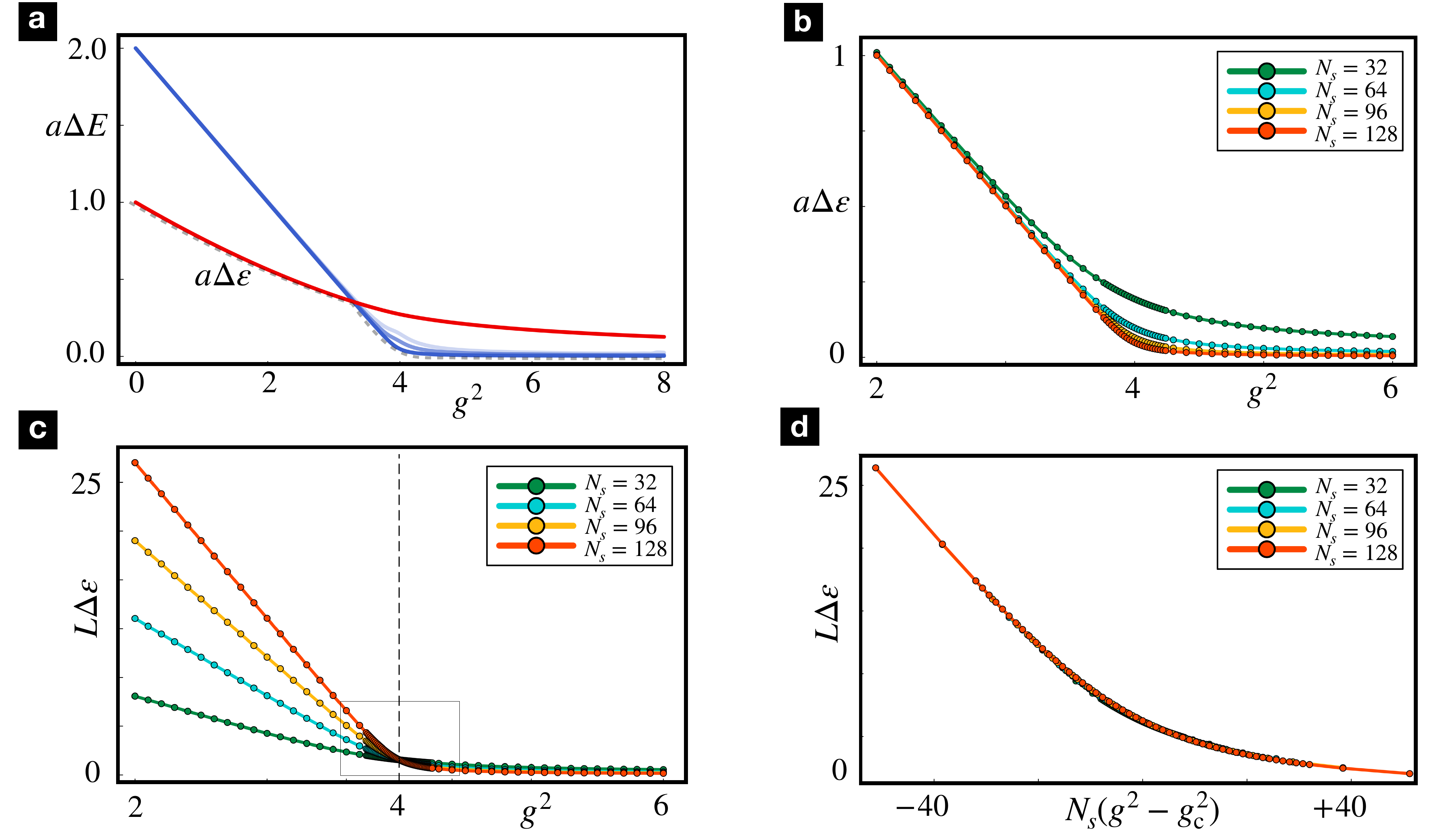}
    \caption{ \textbf{Energy gap along the symmetry line $\boldsymbol{ma=-1}$ at $\mathbf{:\!n_{\rm f}\!:=+1}$ and finite-size scaling around $\mathbf{g^2=4.0}$. }\textbf{(a)} For $N_s=128$ sites: energy difference $\Delta E \equiv E_{p,\boldsymbol{d}}-E_{gs}$ for the first three excitations belonging to the same symmetry sector as the ground state, $\boldsymbol{d}=(1,1,\cdots, 1, 1)$ (blue lines), as well as for the first excitation associated with the promotion of one edge state to the bulk, namely either the sector $\boldsymbol{d}=(1, 1, \cdots, 1, 2, 0)$ or $\boldsymbol{d}=(0, 2, 1, \cdots, 1, 1)$ (red line). Because of a level crossing among the former excitations with the latter ones, the energy gap $\Delta \epsilon$ presents a non-analytic behaviour at $g^2\approx 3.3$. \textbf{(b) -- (d)} Finite-size scaling of the energy gap for $:\!n_{\rm f}\!:=+1$ within the symmetry sector $\boldsymbol{d}\mathbf{=(1,1,\cdots, 1,1)}$: assuming the critical exponents $\nu=1$ and $z=1$ --the latter because the theory is Lorentz invariant-- the finite-size scaling ansatz for the energy gap reads $\Delta \epsilon = L^{-1} f(L |g^2-g_c^2|)$. Thus, we represent: \textbf{(b)} The energy gap $\Delta \epsilon$ as a function of $g^2$ for $L=aN_s=\{32,\,64,\,96,\,128\}$ lengths ($a=1$). \textbf{(c)} $L\Delta \epsilon$ as a function of $g^2$ for several lengths: the different curves cross at $g^2_c=4.0$, which coincides with the critical point for the zero-density regime. \textbf{(d)} These lines merge when plotted versus $L |g^2-g^2_c|$, which reflects the correct choice of the critical exponents, as well as the universal behaviour of the transition. This scaling, together with panel \textbf{(a)}, evidences a quantum phase transition between a gapped SPT phase and a gapless solitonic phase at $g^2_c=4.0$.}
    \label{fig:FSS_deltaE}
\end{figure*}

Let us now discuss one key difference between the solitonic and edge inhomogeneities. Whereas in the latter we have already discussed that there is an accumulation (depletion) of fermions restricted to the edges of the system for $:n_{\rm f}:=+1$ ($:n_{\rm f}:=-1$), this trend is completely different in the solitonic phase. As can be seen in the lower panel of Fig.~\ref{fig:pi_n+1}{\bf (c)}, the rung densities $R_n$ show that the extra fermion (hole) is localised around the position of the anti-kink, which is reminiscent of the Jackiw-Rebbi mechanism of charge fractionalisation~\cite{PhysRevD.13.3398,PhysRevLett.47.986,BELL19831}. This mechanism  occurs in non-linear Yukawa-type QFTs with spontaneous symmetry breaking, allowing for solitonic excitations that connect the different possible vacua. In this case, the discretised models do not have conserved charges, and a single fermion can trigger the formation of a kink-anti-kink pair, each of which confines half of its charge in a quasiparticle. In our case, as we can only have anti-kinks, the whole integer extra (deficit) charge will be localised around the anti-kink (kink).  

We would like to note that, also in contrast to the Jackiw-Rebbi model in which kinks-anti-kinks  appear as stable  excitations of a scalar field, and their width is determined by the inverse of the scalar-field mass, our solitons appear in the ground state of the GNW model upon doping, and tend to maximise their width over much larger length-scales. This is a consequence of the minimisation of the energy associated with the inhomogeneities, and the fact that the $\pi(x)$ are auxiliary fields that do not have their own bare dynamics or mass. Therefore, both the solitons and extra charge displayed in Fig.~\ref{fig:pi_n+1} \textbf{(c)} extend to wider regions. Concerning their profiles, we show in Fig.\ref{fig:rung_density_fits}{\bf (a)} that the rung density distributions for a single-fermion doping in the anti-kink inhomogeneous phase  $g^2> 4$ are all well described in the bulk by
\beq
\label{eq:Rn_fit}
:\!R_n\!:\!\!a=A_{g^2}\cos^2\big(k_{g^2}(n-n_{0})a\big),
\eeq
where $A_{g^2}$ ($k_{g^2}$) describe the amplitude (wavevector) of the density modulation around the position of the soliton $n_{0}=(N_{\rm s}+1)/2$. By plugging Eq.~\eqref{eq:Rn_fit} into Eq.~\eqref{eq:conserved_charges_pi_condensate} and approximating the summations as integrals, one can also get an expression for the pseudoscalar condensate profile
\beq
\label{eq:pi_fit}
\pi_n a=-A_{g^2}\left[ (n-n_0) + \frac{\sin[2k_{g^2}(n-n_0)a]}{2k_{g^2}a}\right]\,,
\eeq
which also agrees with the data, as shown in Fig.\ref{fig:rung_density_fits}\textbf{(b)}. We find that both the amplitude and the wavevector change with the interaction strength $g^2$, as displayed in Fig.~\ref{fig:rung_density_fits}{\bf (c)}. In the region $g^2<4$, the amplitude will scale to zero as the system size increases, whereas it will increase with some power law for $g^2>4$. As the interactions increase, we see that the modulation wave-vector also tends to $k_{g^2}=\pi/aN_{\rm s}$, which is consistent with the charge being confined inside a wide anti-kink. Right at the 
half-filling critical point  $g^2=4$ (Fig.~\ref{fig:scheme_phases}), we see in Fig.~\ref{fig:pi_n+1}{\bf (b)} that the profile of the pseudoscalar condensate is still inhomogeneous, vanishing at the single maximally-symmetric point regarding parity. The key aspect is that the slope of the fermion condensate changes at this point, allowing for the interpolation between the parity-broken asymptotic values to transition from concave to convex, and thus from the boundary localisation to solitonic behaviour.

Before closing this section, let us comment on another qualitative difference between these two phases regarding their entanglement~\cite{RevModPhys.80.517,LAFLORENCIE20161}, a key quantity at the root of the differences between the classical and quantum worlds. To compute the entanglement spectrum~\cite{PhysRevLett.101.010504}, we first split the system into two subsystems, $L$ and $R$, and express the ground state as $\ket{\Psi}=\sum_q\lambda_q\ket{\psi_q}_{\rm L}\otimes\ket{\psi_{q}}_{\rm R}$, where $\lambda_q\in[0,1]$ represent the so-called Schmidt coefficients. The entanglement spectrum is then defined as the logarithmic scale of these coefficients, $\epsilon_q=-2\log\lambda_q$ 
	, which can be directly obtained from the MPS simulations, as they give direct access to the reduced density matrices of any specific bipartition as one performs the corresponding variational sweeps. As highlighted in~\cite{PhysRevB.81.064439,PhysRevLett.104.130502},  degeneracy in the entanglement spectrum is indicative of SPT phases. This degeneracy remains intact under symmetric perturbations as long as the many-body gap stays open.

In Fig.~\ref{fig:ES_N+1}, we present our MPS numerical results on entanglement spectra for a system with $N_{\rm s}=128$ lattice sites. Starting from $g^2=2$ (see Fig.~\ref{fig:ES_N+1} {\bf (a)}), the entanglement spectrum shows a clear two-fold degeneracy associated with the topological nature of this phase. This degeneracy is the manifestation of the physics of the zero-energy edge states, which at the level of the reduced density matrix also appear at the bipartition that separates the system in two blocks~\cite{PhysRevLett.104.130502}. We have checked that this degeneracy is robust for all $g^2<4$, whereas it is broken for $g\geq 4$, as we display in Fig.~\ref{fig:ES_N+1}\textbf{(b)} for $g^2=8$. In the following section, we will provide evidence showing that the disappearance at $g^2\geq4$ is caused by the vanishing of the many-body energy gap in the whole solitonic phase.

\subsubsection{Low-energy excitations and  finite-size  scaling  for an extra fermion above half filling}

In the variational MPS formalism, one needs not restrict to the ground state but may also target low-energy excitations. This is done by imposing an orthogonality constraint with the previously found ground state $\ket{\varphi_{\rm gs}}$, and proceeding recursively by enlarging the set of orthogonality conditions as one climbs the energy ladder. In practice, these orthogonality conditions can be implemented in the optimisation process if one instead minimises the energy of the Hamiltonian $H'=H+\sum_i |w_i|\ket{\varphi_i}\bra{\varphi_i}$, where the states $\ket{\varphi_i}$ are the already computed ground and excited states. If the coefficients $|w_i|$ are large enough, the energy penalties favour the target excitation to be the ground state \cite{stoudenmire2012studying}. In our GNW model, there is a slight complication that arises at $ma=-1$ and is related to the extensive number of conserved quantities: there may be level crossings between different symmetry sectors. Even if we did not find any level crossing for the ground state problem with $:\!\! n_{\rm f}\!\!:=\pm 1$ doping, this is not excluded  when considering excitations. In order to account for this, we computed the excited states associated with different symmetry sectors. Following a similar idea as before, this can be done by using the dimer number operators $D_n$, e.g. adding a term of the type $|w|D_n^2$ or $-|w|D_n$ to select a state with $d_n=1$ or $d_n=0$, respectively. Since these operators are functions of the conserved charges, they do not modify the eigenstates of the Hamiltonian, but their energy ordering. Following this strategy, we show in Fig.~\ref{fig:FSS_deltaE}{\bf (a)}, for a system of $N_s=128$ sites with $:\!\!n_{\rm f}\!\!:=+1$, the energy gap $\Delta\epsilon=\min\{\Delta E_{p,\boldsymbol{d}}=E_{p,\boldsymbol{d}}-E_{\rm gs},\forall p,\boldsymbol{d}\}$ as a function of $g^2$, where $p$ labels all possible excitations within the sector $\boldsymbol{d}$. As it can be seen, the energy gap depicted with a black dashed line displays a non-analyticity at $g^2\approx 3.3$, even for finite system sizes, which is ultimately related to a level crossing in the gapped SPT phase. In this figure, we represent in blue the minimal excitation energy for $\boldsymbol{d}\in\mathbb{D}_{\rm gs}$, while the red curve considers excitations in $\boldsymbol{d}\in\{(0,2,1,\cdots, 1, 1,1),\,(1,1,1,\cdots,1,2,0)\}$, which are symmetry sectors in which one of the edge states has been promoted to a double occupancy of the nearest dimer. As can be seen, for weak interactions excitations in this latter sector have lower energies and, thus, are responsible for the energy gap. However, at   $g^2\approx 3.3$, a level crossing takes place, and the low-energy excitations belong again to $\boldsymbol{d}\in\mathbb{D}_{\rm gs}$.

\begin{figure}
    \centering
    \includegraphics[width=0.70\linewidth]{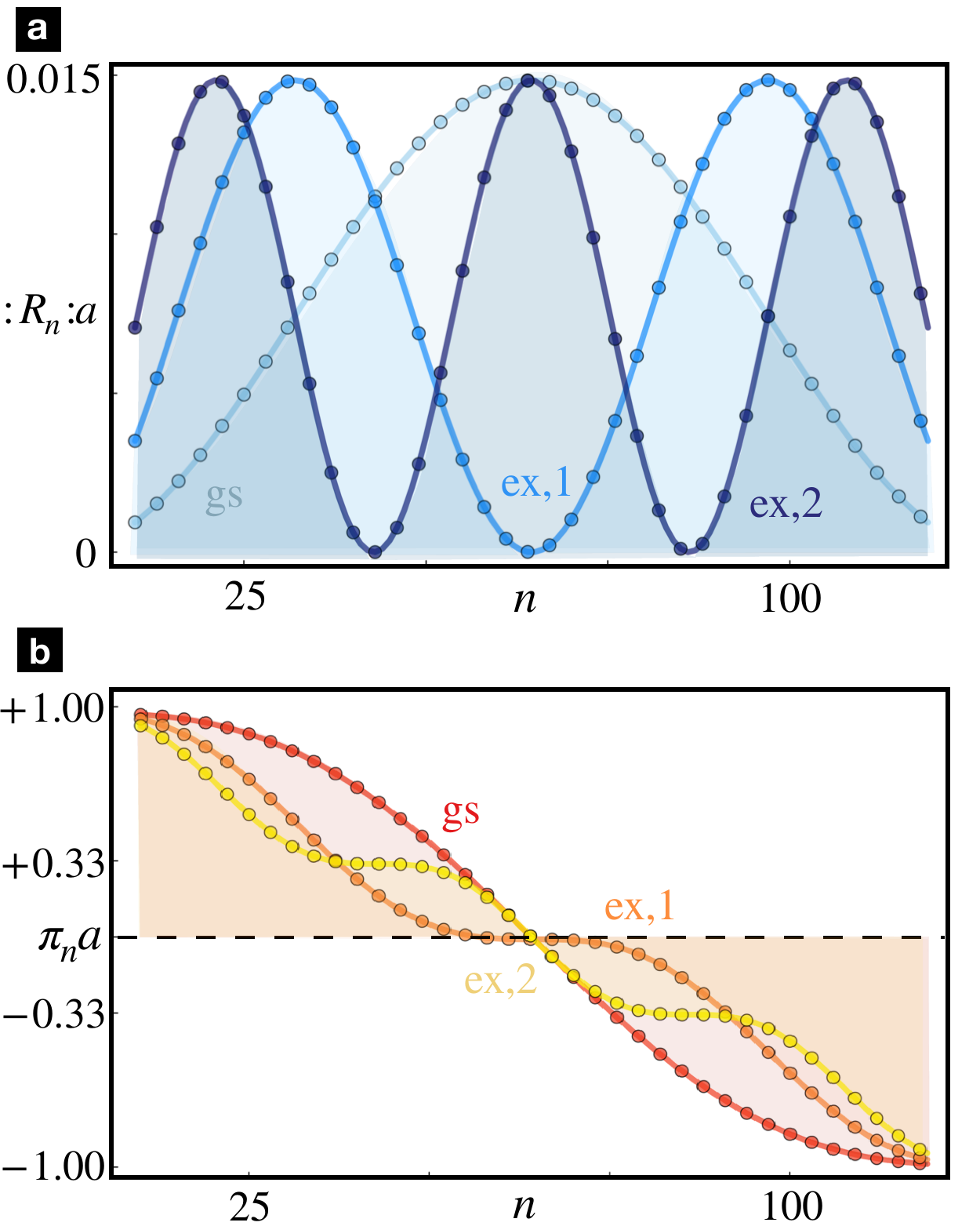}
    \caption{ \textbf{First excited states at $\mathbf{:\!n_{\rm f}\!:=+1}$ and $\mathbf{g^2=8}$.} \textbf{(a)} Fit of the rung density $:\!R_n\!:$ data (points) to Eqs.~\eqref{eq:Rn_fit},~\eqref{eq:Rn_fit_ex} (lines) for the ground state (light blue), first excited state (blue) and second excited state (dark blue) in the bulk, being the fitted parameters for the excited states $(A_{8.0,\, 1}, \;k_{8.0,\, 1}N_sa/\pi, \; \varphi_{1})=(1.484\cdot 10^{-2},\,1.969,\, \pi/2)$, $(A_{8.0,\, 2},\; k_{8.0,\,2},\; \varphi_2)=(1.484\cdot 10^{-2},\, 2.954,\, 0.0)$, so that $A_{8.0}\approx A_{8.0,\,1}\approx A_{8.0,\,2}$ and $k_{8.0}\approx k_{8.0,\,1}/2 \approx k_{8.0,\,2}/3$. \textbf{(b)} Fit of the pseudoscalar condensate data (points) to Eqs.~\eqref{eq:pi_fit},~\eqref{eq:pi_fit_ex} (lines) for the ground state (red), first excited state (orange) and second excited state (yellow) in the bulk, being the fitted parameters for the excited states the ground and first two excited states the same as those obtained for $:\!R_n\!:$.}
    \label{fig:EN_N+1}
\end{figure}

Apart from the level crossing, we note that the minimal energy gap $\Delta\epsilon$ vanishes for $g^2 \geq 4$ in the thermodynamic limit. Indeed, a finite-size scaling of this gap, depicted in Fig.~\ref{fig:FSS_deltaE}\textbf{(b)--(d)}, shows that it has a non-analytic behaviour at the critical point $g_c^2=4$, coinciding with the one for $:\!\!n_{\rm f}\!\!:=0$. However, in contrast to the scaling behaviour of the half-filled limit, which is completely captured by an Ising critical point, here the transition is between a gapped (SPT) and a gapless (solitonic) phase, so we expect that a careful finite size scaling should contain new physics.

Finally, concerning the $:\!R_n\!:$ and $\pi_n$ profiles for the lowest-lying excitations in the solitonic phase, we show in Fig.~\ref{fig:EN_N+1} \textbf{(a)--(b)} that they are captured by expressions analogous to Eq.~\eqref{eq:Rn_fit} and ~\eqref{eq:pi_fit}, namely 
\beq
\label{eq:Rn_fit_ex}
:R^{\rm ex,q}_n:\!\!a=A_{g^2,q}\cos^2\big(k_{g^2,q}(n-n_{0})a+\varphi_q\big),
\eeq
for the excited rung densities, and
\beq
\label{eq:pi_fit_ex}
\pi^{\rm ex,q}_n\, a=-A_{g^2,q}\left[ (n-n_0) + \frac{\sin[2k_{g^2,q}(n-n_0)a+\varphi_q]}{2k_{g^2,q}a}\right]\,,
\eeq
for the pseudoscalar profile. There is thus a
 similar modulation, but different wavevectors and phases. In particular, we find that $\varphi_1\approx\pi/2$ and $k_{g^2,1}=2\pi/aN_{\rm s}$ for the first excitation, and $\varphi_2\approx0$ and $k_{g^2,2}=3\pi/aN_{\rm s}$ for the second,  capturing the structure of the low-energy solitons in the figure.

\subsubsection{Real-space fragmentation for two extra fermions above half filling}\label{sec:fragmentation}

\begin{figure*}
    \centering
    \includegraphics[width=1\linewidth]{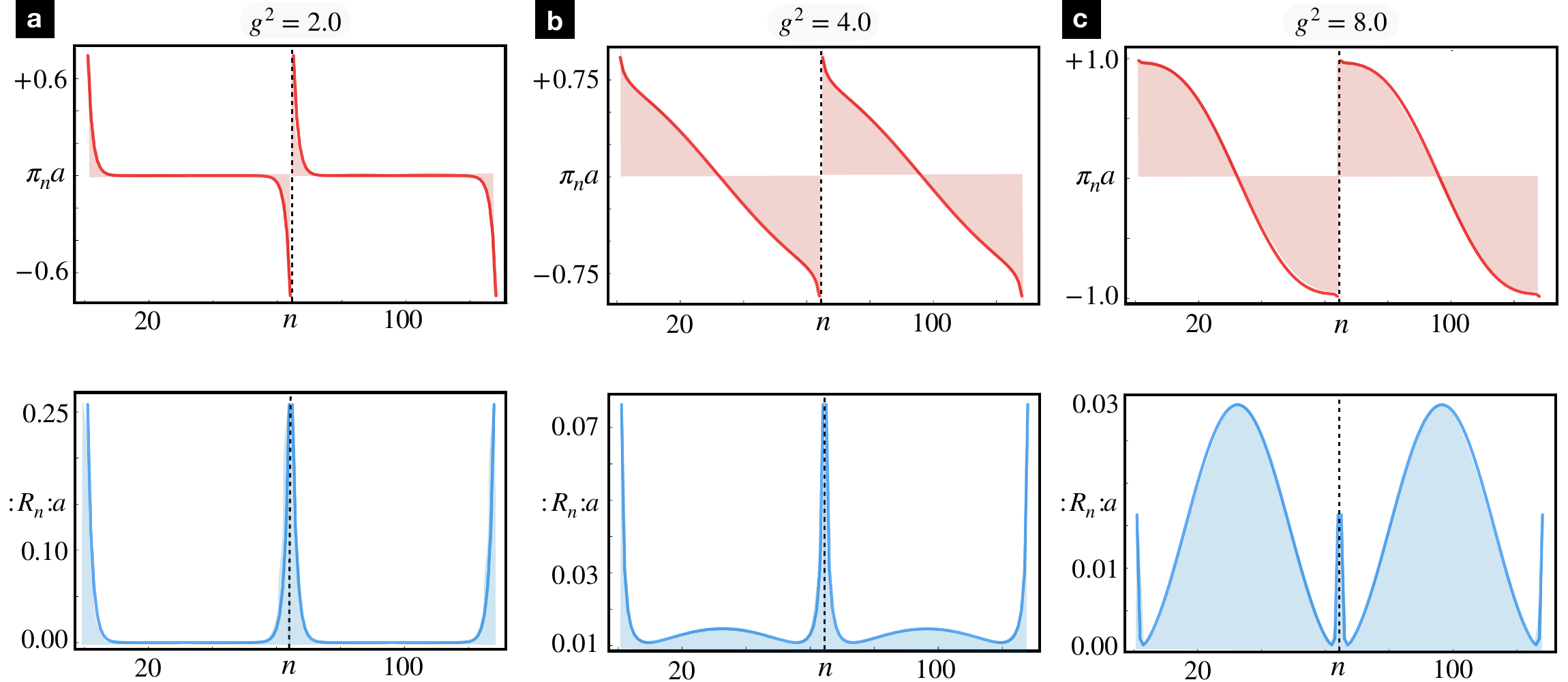}
    \caption{ \textbf{Inhomogeneities along the symmetry line $\boldsymbol{ma=-1}$ at $\mathbf{:\!n_{\rm f}\!:=+2}$.} Pseudoscalar condensate $\pi_n$ (upper panels) and rung density $:\!R_n\!:$ (lower panels) for two fermion doping at: \textbf{(a)} $g^2=2.0$, \textbf{(b)} $g^2=4.0$ and \textbf{(c)} $g^2=8.0$. The second extra fermion induces a chain fragmentation, such that the profiles consist of two copies of the single doped ones depicted in Fig.~\ref{fig:pi_n+1}. Among all possible chain fragmentations, the true ground state configuration is the one where the fragmentation takes place at the middle of the chain.}
    \label{fig:pi_n+2}
\end{figure*}

\begin{figure}
    \centering
    \includegraphics[width=0.8\linewidth]{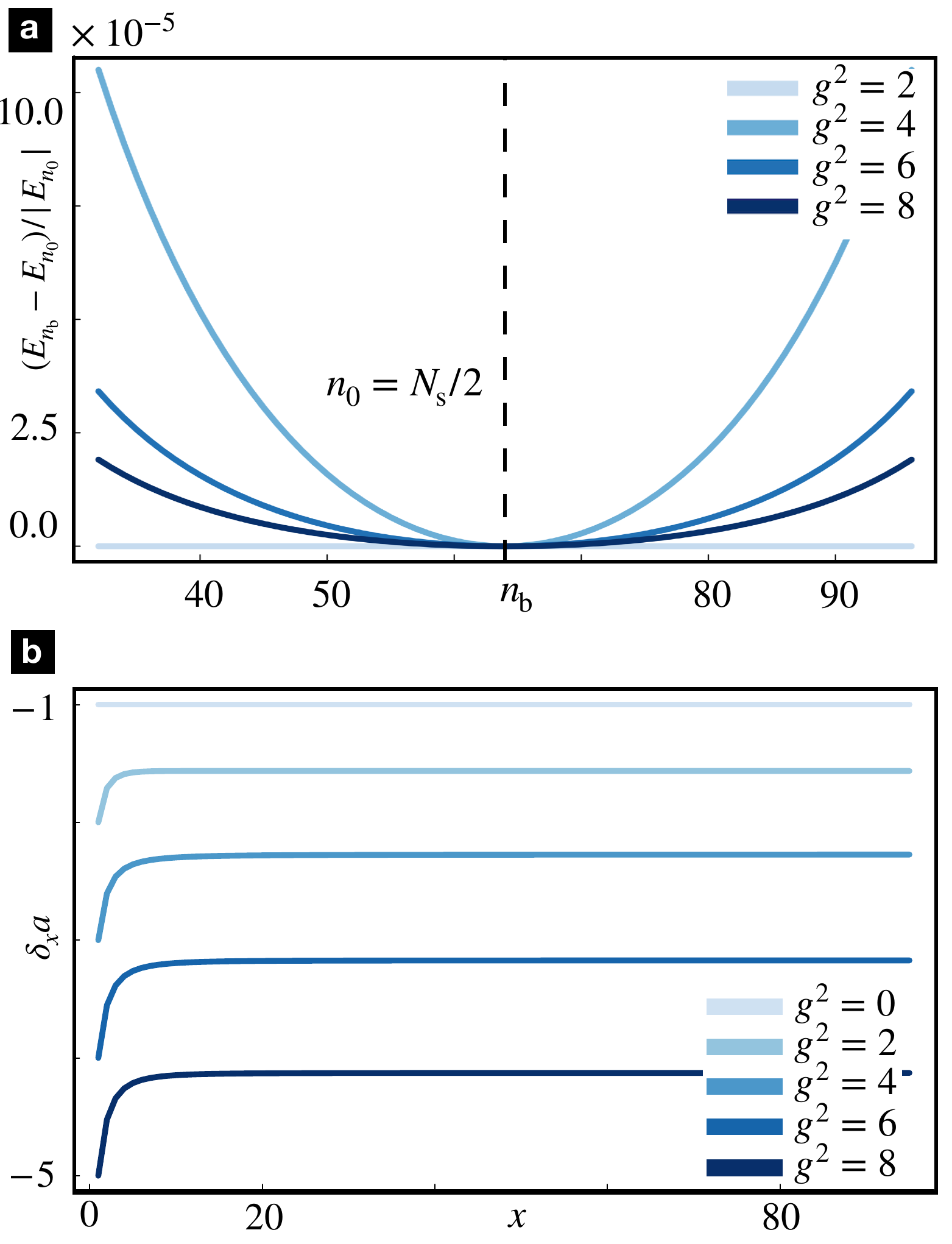}
    \caption{\textbf{Ground state energy in a fragmented chain.} \textbf{(a)} Relative difference of the ground state energy for a chain with two doped fermions as a function of the fragmentation point $n_{\rm b}$. The configuration with the lowest energy is the one with $n_{\rm b}=N_s/2$, rising as $n_{\rm b}$ is moved to the boundaries. The orders of magnitude of the relative difference turn out to be $10^{-5}$ for $g^2 \gtrsim 4$, while it is around $10^{-10}$ for $g^2 \lesssim 2$. This very small relative difference in energies makes extremely challenging the task of finding the true ground state without appealing to strategies based on the conserved quantities and/or chain fragmentation. \textbf{(b)} ground state energy difference $\delta_{x}\equiv E(x+1)-E(x)$ for a single-doped chain as a function of the total number of sites $x$. For $g^2>0$ it is a monotonously increasing function, reaching a $g^2-$dependent maximum for a sufficiently large $x$. Along these nearly flat intervals, the cost of moving the fragmentation point $n_{\rm b}$ is small.} 
    \label{fig:fragmentation_location}
\end{figure}

Let us now increase the doping to two fermions/holes, focusing first on the $:\!\!n_{\rm f}\!\!:=+ 2$ scenario. In this case, the accommodation of the extra fermion requires abandoning the sector $\boldsymbol{d}=(1,1,\cdots,1,1)$. The only possibility is that one of the conserved charges at a site $n_{\rm b}$ changes from $d_{n_{\rm b}}=1\to d_{n_{\rm b}}=2$, such that one of the bulk dimers gets maximally populated $\boldsymbol{d}_{\rm b}=(1,1,\cdots,1,2,1,\cdots,1,1)$. It is then a simple matter to see that the tunnelling term in Eq.~\eqref{eq:rung_hamiltonian} is no longer effective across the $n_{\rm b}$-th diagonal link due to the Pauli exclusion principle (see Fig.~\ref{fig:scheme_dimerised} \textbf{(a)}). Since there is no other term in the Hamiltonian~\eqref{eq:rung_hamiltonian} that couples neighbouring rung orbitals, one can see that the Hamiltonian is broken into two commuting pieces, separated by the breaking site $n_{\rm b}\in\mathbb{Z}_{N_s-1}$, which is a direct real-space manifestation of the Hilbert-space fragmentation~\cite{Moudgalya_2022} discussed previously. Although the exact localisation of the fragmentation is yet to be determined, we know that it will remain immobile and thus partition the system into two sub-chains, introducing an effective interface. This provides another mechanism of real-space fragmentation alternative to the blockade-separated regions identified in~\cite{PhysRevLett.134.010411}. 

If we define $P_{n_{\rm b}}$ as the projector onto the corresponding sector $\mathcal{F}_{N_{\rm s}+2}^{\boldsymbol{d}_{\rm b}}$, the Hamiltonian breaks into two  parts $
P_{n_{\rm b}}HP_{n_{\rm b}}=H_{\rm L}^{n_{\rm b}}+H_{\rm R}^{n_{\rm b}}
$ with
\beq
\begin{split}
\vspace{-2cm}H_{\rm L}^{n_{\rm b}}\!&=\!\!\sum_{n=1}^{n_{\rm b}-1}\!\!\left(\frac{\ii}{a} \Phi^{{\dagger}}_{n,+}\Phi^{\phantom{\dagger}}_{n+1,-}+{\rm H.c.}\right)+\!\!\sum_{n=1}^{n_{\rm b}}\!\!\frac{g^2}{a}\!N_{n,+}N_{n,-}-\\
& -\sum_{n=1}^{n_{\rm b}}\!\!\frac{g^2}{2a}\left(N_{n,+}+N_{n,-}\right)=: H(1,n_{\rm b})\,, \\
H_{\rm R}^{n_{\rm b}}\!\!\!&=\!\!\!\!\!\sum_{n=n_{\rm b}+1}^{N_{\rm s}-1}\!\!\!\left(\frac{\ii}{a} \Phi^{{\dagger}}_{n,+}\Phi^{\phantom{\dagger}}_{n+1,-}+{\rm H.c.}\!\right)\!+\!\!\!\sum_{n=n_{\rm b}+1}^{N_{\rm s}}\!\!\frac{g^2}{a}\!N_{n,+}N_{n,-}-\\
& -\sum_{n=n_{\rm b}+1}^{N_{\rm s}}\!\!\frac{g^2}{2a}\left(N_{n,+}+N_{n,-}\right)=: H(n_{\rm b}+1, N_{\rm s})\,, \vspace{-2cm}\\
\end{split}
\eeq
where we  introduced the number operators $N_{n,\pm}=\Phi^{{\dagger}}_{n,\pm}\Phi^{\phantom{\dagger}}_{n,\pm}$, and $H(n_1, n_2)$ denotes the original Hamiltonian \eqref{eq:rung_hamiltonian} in a chain with sites labelled from $n_1$ to $n_2$.

This fragmentation has a direct consequence on the inhomogeneities and the nature of the phases for $g^2<4$ and $g^2>4$. In the non-interacting limit $g^2=0$, this is easy to understand as the doping-induced fragmentation changes the boundary conditions around site $n_{\rm b}$, which can indeed be understood as a defect. According to our understanding of SPT defects~\cite{RevModPhys.88.035005}, we expect that new bound states will be localised around the defect, which will host the extra doped fermion. It is interesting to note that, algebraically, the 2-fermion populated link $
\Phi_{n_{\rm b},+}^\dagger\Phi_{n_{\rm b}+1,-}^\dagger\ket{0}
$ can be formally understood as a left- and a right-localised topological edge modes, in full analogy to the edge states that appear at the system boundaries~\eqref{eq:edge_oparators}, but now localised around the defect at $n_{\rm b}$. We remark that this defect is the result of doping and quantum statistics, contrary to the standard situation in which one considers an external local perturbation in the system that breaks translational invariance explicitly. In fact, its explicit position $n_{\rm b}$ is not fixed externally, but will result from energetic considerations.

We now use our MPS approach to explore how this phenomenon can be extended to interacting regimes, which will also allow us to determine in which spatial point $n_{\rm b}$ the defect appears, and also corroborate that the ultra-local localisation of the extra charge changes into an exponential one when turning on the interactions. In Fig.~\ref{fig:pi_n+2}{\bf (a)}, we represent our results for $g^2=2$, which clearly show that the fragmentation occurs right at the centre of the chain $n_{\rm b}=N_{\rm s}/2$. The profile of the pseudoscalar condensate (see upper panel) is essentially that of two consecutive copies of the single-fermion doping case discussed in Fig.~\ref{fig:pi_n+1}{\bf (a)}. Once again, the distribution is an odd function $\pi_n=-\pi_{N_{\rm s}+1-n}$, such that parity symmetry is also restored at this filling. Bringing our attention to the rung densities on the lower panel of Fig.~\ref{fig:pi_n+2}{\bf (a)}, we also corroborate that the topological defect hosts the extra doped fermion, which is exponentially distributed to the left and right of $n_{\rm b}$, mimicking the charge distribution of the topological edge states. Let us note that for hole doping $:\!n_{\rm f}\!:=-2$, the discussion is completely analogous but we instead would find the chirally-reversed configuration, such that the extra holes are confined to the boundaries and the central defect.

We now explore the strongly-interacting regime $g^2=8$, where we recall that parity gets spontaneously broken at half filling. In this case, using the constraints~\eqref{eq:conserved_charges} for the double occupied link $d_{n_{\rm b}}=2$, and taking into account that the extra doped charge can only lead to rung occupancies $R_{n_{\rm b}}+R_{n_{\rm b}+1}\leq 3/a$, we see that the previous monotonously-decreasing condition turns into an increasing one across this point. Revisiting Eq.~\eqref{eq:monotonous}, we can now say
\beq
\label{eq:monotonous_N+2}
\begin{split}
:\!n_{\rm f}\!:=+2,\hspace{2ex} \pi_n>\pi_{n+1},\forall n\neq n_{\rm b},\,\,{\rm but}\,\, \pi_{n_{\rm b}}<\pi_{n_{\rm b}+1},\\
:\!n_{\rm f}\!:=-2,\hspace{2ex} \pi_n<\pi_{n+1},\forall n\neq n_{\rm b},\,\,{\rm but}\,\, \pi_{n_{\rm b}}>\pi_{n_{\rm b}+1}.
\end{split}
\eeq
We can thus see that the pseudoscalar condensate can actually increase (decrease) abruptly right at the defect caused by $d_{n_{\rm b}}=2$ ($d_{n_{\rm b}}=0$). This is confirmed by our MPS results in Fig.~\ref{fig:pi_n+2}{\bf (c)}, which show how the fermion condensate increases right at the $n_{\rm b}=N_{\rm s}/2$, position where the chain fragments. Therefore, we see that for $:\!n_{\rm f}\!:=+2$ the pseudoscalar condensate develops two consecutive anti-kinks such that the total topological charge is now $Q_\pi^{:n_{\rm f}:=+2}=-2$. As already noted in the previous section, we do not find an interpolation of kinks and anti-kinks due to the constraints of the conserved charges, but instead a succession of solitons with the same charge. Finally, looking into Fig.~\ref{fig:pi_n+2}{\bf (c)}, we see that each of these anti-kinks confines one of the doped fermions, which now get localised around each of the regions in which the fermion condensate vanishes, $n_1=N_{\rm s}/4$ and $n_1=3N_{\rm s}/4$, instead of the system edges and the fragmentation point $n_{\rm b}=N_{\rm s}/2$.

To further support the physical interpretation of the fragmentation point $n_{\rm b}$, we now calculate the ground-state energy obtained by pasting two optimised MPS segments with one extra fermion each, for a fixed total number of sites $N_{\rm s}=128$. This approach allows us to vary the value of $n_{\rm b}$ and access to the different symmetry sectors easily. As shown in Fig.~\ref{fig:fragmentation_location}, the MPS energy landscape exhibits a shallow dependence on the fragmentation location, with extremely small relative energy differences for both weak and strong couplings. These results indicate that there exists a manifold of nearly degenerate configurations, each corresponding to a different internal positioning of the domain wall.
This dense clustering of quasi-degenerate states illustrates the practical difficulty of identifying the true ground state via direct MPS optimisation, especially in the absence of guiding conserved quantities or a priori knowledge of the fragmentation pattern. 
\begin{figure*}
    \centering
    \includegraphics[width=1.0\linewidth]{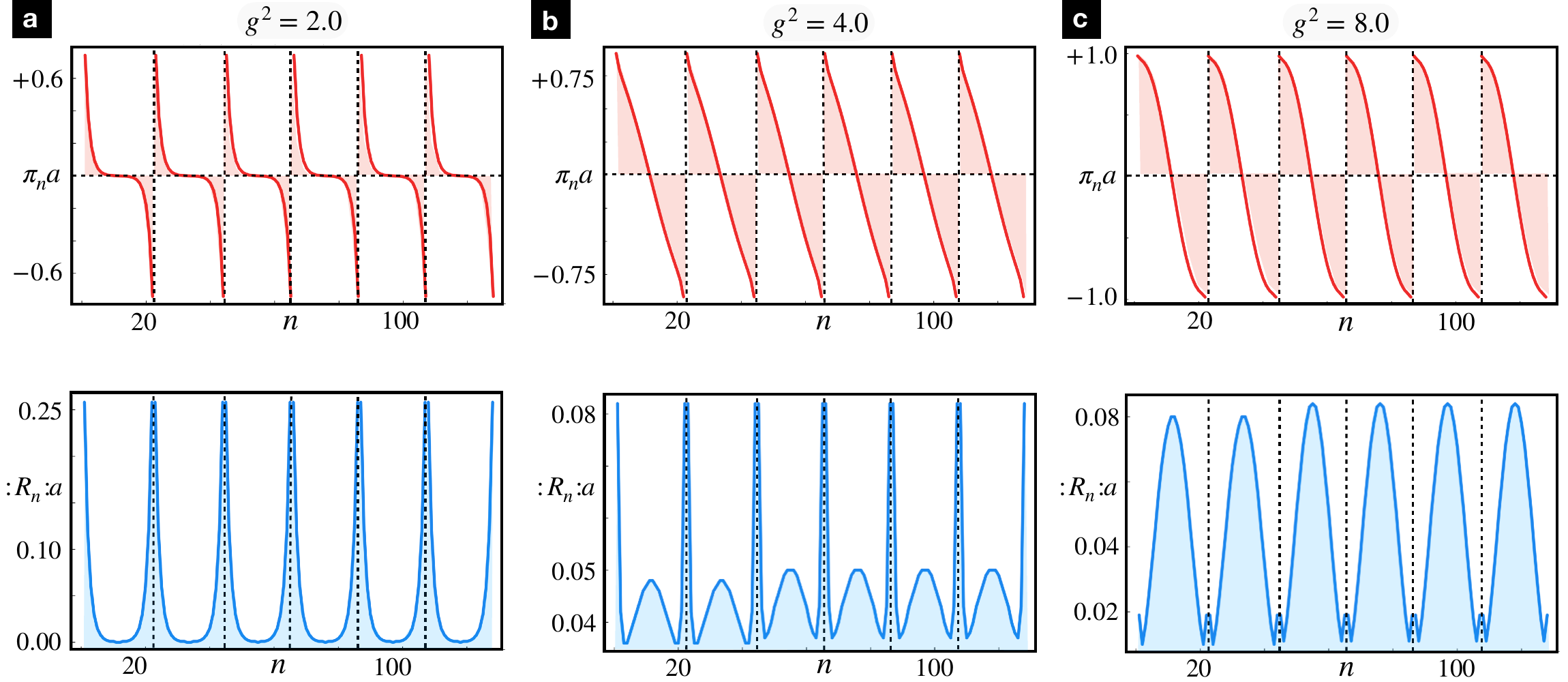}
    \caption{\textbf{Crystals along the symmetry line $\boldsymbol{ma=-1}$ at low densities.} Pseudoscalar condensate $\pi_n$ (upper panels) and rung density $:\!R_n\!:$ (lower panels) for $:\!n_{\rm f}\!:=6$ at: \textbf{(a)} $g^2=2.0$, \textbf{(b)} $g^2=4.0$ and \textbf{(c)} $g^2=8.0$. As we increase the fermion/hole doping, the chain is broken into $|\!\!:\! n_{\rm f}\!:\!\!|$ independent subchains, each of which carries $|\!\!:\! n_{\rm f}\!:\!\!|=+1$. It is noteworthy to outline that, since the filling is not commensurate, not all the subchains have the same length --in this case the first two possesses one more site--, so there is a microscopic degeneracy on the spectrum consisting of how they are arranged.}    \label{fig:commensurate_crystals}
\end{figure*}

It is noteworthy to outline an alternative approach to understanding this quasi-degeneracy in terms of the energy of a single chain. Due to the chain fragmentation, the total energy for a chain with $|\!\!:\!\!n_{\rm f}\!\!:\!\!|=2$ corresponds to the sum of the ones of the two subchains, that is, $E_{\rm tot}=E(N_1)+E(N_2)$, where $E(N_s)$ stands for the energy of a single-doped chain with $N_{\rm s}$ sites. Thus, the aforementioned quasi-degeneracy can be understood in terms of the energy difference $\delta_{x} := E(x+1)-E(x)$, depicted in Fig.~\ref{fig:fragmentation_location} \textbf{(b)} for several values of $g^2$. As it can be seen, $\delta_{x}$ saturates for a sufficiently large number of sites, except for $g^2=0$, which is completely constant since the addition of an extra site corresponds to simply including an independent dimer. This saturation implies that $E_{\rm tot}=E(N_{\rm s}/2)+E(N_{\rm s}/2) = E(N_{\rm s}/2+1)+E(N_{\rm s}/2-1) -\delta_{N_{\rm s}/2}+ \delta_{N_{\rm s}/2-1}\approx E(N_{\rm s}/2+1)+E(N_{\rm s}/2-1)$. More generically, for a small $\Delta$ we would have that $E_{\rm tot}=E(N_{\rm s}/2)+E(N_{\rm s}/2) \approx E(N_{\rm s}/2+\Delta)+E(N_{\rm s}/2-\Delta)$. Of course, for $g^2>0$ the function $\delta_x$ is not completely flat, but it is a monotonously increasing function so that the energy decrease coming from enlarging one subchain is lower than the energy increase from shrinking the other one. This difference is significant when taking either $N_1$ or $N_2$ relatively small, which disfavours highly-asymmetric fragmented chains.

\subsubsection{Topological crystals and soliton lattices for many fermions above half filling}

 \begin{figure*}
    \centering
    \includegraphics[width=0.8\linewidth]{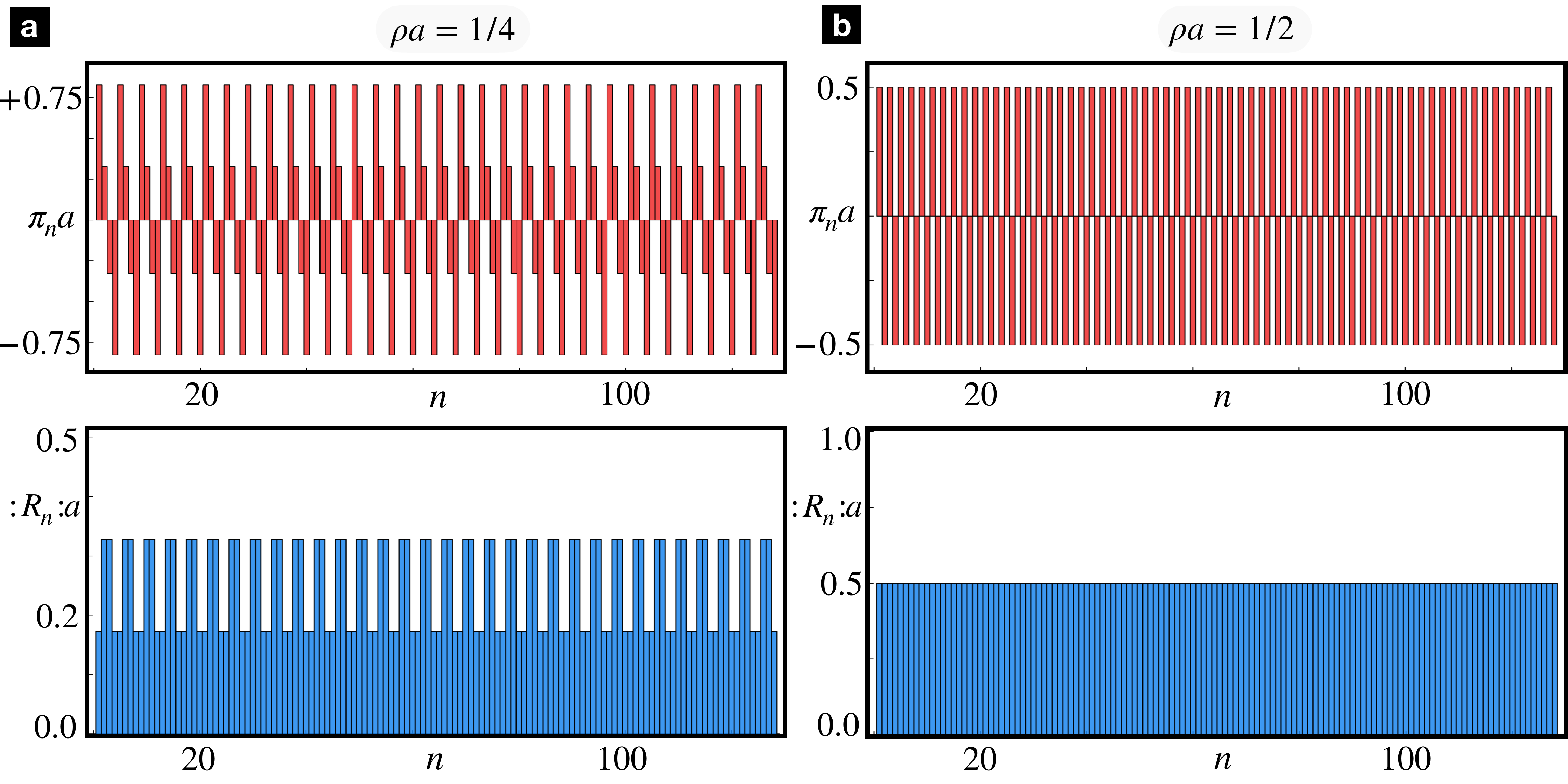}
    \caption{\textbf{Crystals along the symmetry line $\boldsymbol{ma=-1}$ at high commensurate densities and $\mathbf{g^2=8.0}$.} \textbf{(a)} At $\rho a=1/4$, there are still inhomogeneities for both the pseudoscalar condensate $\pi_n$ and the rung densities $:\!R_n\!:$. \textbf{(b)} On the contrary, $:\!R_n\!:$ becomes homogeneous at $\rho a=1/2$, since it is an even function inside each subchain and for this filling each extra particle is distributed in only two sites.}
    \label{fig:commensurate_crystals_high}
\end{figure*}

The results in Fig.~\ref{fig:commensurate_crystals} reinforce the central message of our work: {the emergence of crystalline structures through Hilbert-space fragmentation in the doped GNW model}. As shown in Fig.~\ref{fig:commensurate_crystals}, as one increases the doping but still at low densities, e.g. $:\!\!n_{\rm f}\!\!: = 6$ extra fermions in a chain with $N_s=128$ sites, the pseudoscalar condensate $\pi_n$ and the local rung density $:\!R_n\!:$ reveal again a fragmentation into independent subchains. For $g^2=2$ (left panels),  these subchains are connected by topological defects, each of which can bind one of the doped fermions into exponentially localised zero modes. Importantly, for non-commensurate fillings, the subchains cannot all have equal length --- in the configuration shown, the first two segments are longer by one site --- leading to a {microscopic degeneracy} associated with how the asymmetry is distributed along the chain. This degeneracy, already anticipated in the energy profile of the two-soliton sector (Fig.~\ref{fig:fragmentation_location} \textbf{(a)}), is now seen to persist and compound as more solitons are added.

In contrast, the right panels of Fig.~\ref{fig:commensurate_crystals} illustrate the case of stronger interactions $g^2=8$, for which the pseudoscalar condensate develops a solitonic profile for each individual subchain, leading to a sequential arrangement of anti-kinks topological charges $Q_\pi=-1$. In this case, the $:\!n_{\rm f}\!: = +6$ extra doped fermions are no longer localised at the interface between fragmented subchains, but instead in the respective bulk around each of the solitons.   

We now explore the regime of {higher commensurate fillings}, where the system enters more robust, gapped phases. As shown in Fig.~\ref{fig:commensurate_crystals_high}, at $\rho = 1/4a$ fermion densities, the pseudoscalar condensate $\pi_n$ and the rung density $:\!R_n\!:$ still exhibit visible modulations with a four-site unit cell. At $\rho = 1/2a$, the ground state develops {a homogeneous rung density} that signals a locking of the crystalline condensate pattern to that of the lattice structure, such that each solitonic excitation spreads only across two sites, forming a {regular dimerised array of tightly bound topological defects}. The fact that these states minimise the energy and form clear plateaux in the compressibility,  as shown in the finite-$\mu$ simulations in Sec.~\ref{Grand-canonical}, supports their identification as gapped {solitonic crystals} stabilised by a Hilbert space fragmentation pattern that is exactly --or approximately-- commensurate with the original chain.

\begin{figure*}
    \centering
    \includegraphics[width=1.0\linewidth]{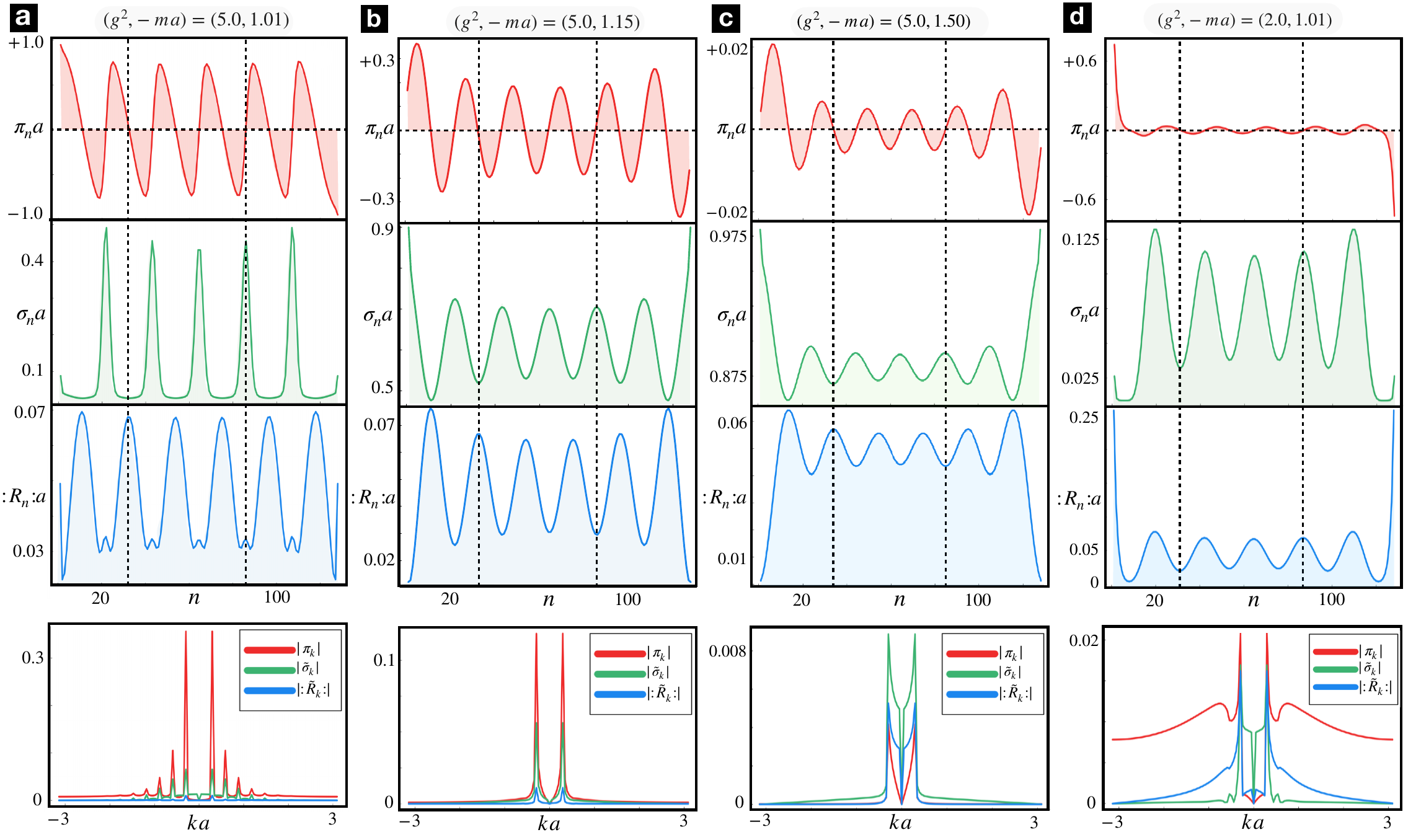}
    \caption{ \textbf{Inhomogeneities out of the symmetry line $\boldsymbol{ma=-1}$.} From top to bottom: pseudoscalar condensate $\pi_n$ (red), scalar condensate $\sigma_n$ (green) and rung density $:\!R_n\!:$ (blue) and their amplitude spectra $|\pi_k|$, $|\tilde{\sigma}_k|$, $|\!\!:\!\tilde{R}_k\!:\!\!|$ --where the tilde denotes a previous subtraction of their mean-- for $:\!n_{\rm f}\!:=6$ and \textbf{(a)} $(g^2, -ma)=(5.0,\,1.01)$, \textbf{(b)} $(g^2, -ma)=(5.0,\,1.15)$, \textbf{(c)} $(g^2, -ma)=(5.0,\,1.50)$, \textbf{(d)} $(g^2, -ma)=(2.0,\,1.01)$. \textbf{(a)} As we move slightly away from the $ma=-1$ line at $g^2=5$, the discontinuities shown in the pseudoscalar condensate become smoother, and coincide with peaks of the scalar condensate. \textbf{(b)} Sufficiently far away from the symmetry line, the condensates present a sine/cosine-like behaviour with a wavevector $k=2\pi\rho$. Likewise, following the intuition developed for the $ma=-1$ regime, the extra charge tends to accumulate around the negative-slope regions of the pseudoscalar condensate. \textbf{(c)} If we move further away from the symmetry line, the inhomogeneities remain in shape but with significantly suppressed amplitudes. \textbf{(d)} As opposed to panel \textbf{(a)}, inside the region associated with the SPT at zero density the discontinuities are abruptly lost when moving away from the $ma=-1$ line, developing the three quantities oscillations with also very small amplitudes in the bulk.} 
    \label{fig:inhomogeneities_out_of_ma-1}
\end{figure*}

The appearance of both compressible and incompressible phases seen in Figs.~\ref{fig:mu_phases}--\ref{fig:rho_kappa_gsq} for $\mu>0$ can be understood again in terms of the chain fragmentation. As explained throughout this section, doping the system with $|\!\!:\!\!n_{\rm f}\!\!:\!\!|>1$ fermions/holes implies that the chain is broken into $|\!\!:\!\!n_{\rm f}\!\!:\!\!|$ subchains, each one containing one fermion/hole. Denoting $N_i$ as the number of sites of the $i$-th subchain, the set of $\{N_i\}_{i=1}^{|:n_{\rm f}:|}$ fulfills the condition $\sum_{i=1}^{|:n_{\rm f}:|} N_i=N_{\rm s}$, and the total grand-canonical energy reads $E=\sum_{i=1}^{|:n_{\rm f}:|} E(N_i) - \mu (N_{\rm s}+|\!\!:\!\!n_{\rm f}\!\!:\!\!|)$. In order to analyse the existence of these two phases, it is necessary to take into account that, concerning the possible values of $N_i$, the most-favoured energetic configuration is the one that minimises the difference among the lengths of all subchains. This follows from the assumption that $\delta_{x}$ is monotonously increasing, as indicated in Fig.~\ref{fig:fragmentation_location} \textbf{(b)} for $g^2\neq 0$, and it is also supported by the results shown in Figs. \ref{fig:pi_n+2}, ~\ref{fig:commensurate_crystals} and ~\ref{fig:commensurate_crystals_high}. Thus, from Figs.~\ref{fig:mu_phases} and \ref{fig:rho_kappa_gsq}, we observe that the compressible phases take place only at low densities, which leads to relatively large values for the $N_i$'s, falling consequently in the nearly-flat region of $\delta_{x}$ of Fig.~\ref{fig:fragmentation_location} \textbf{(b)}. If we compute the value of $\mu$ at which the ground state energy associated with the $|\!\!:\!\!n^{(1)}_{\rm f}\!\!\!:\!\!|$ sector equals the one with doping $|\!\!:\!\!n^{(2)}_{\rm f}\!\!\!:\!\!|>|\!\!:\!\!n^{(1)}_{\rm f}\!\!\!:\!\!|$, we would get
\beq
\mu = \frac{\sum_{i=1}^{|:n^{(2)}_{\rm f}:|} E(N^{(2)}_i)-\sum_{i=1}^{|:n^{(1)}_{\rm f}:|} E(N^{(1)}_i)}{|\!\!:\!\!n^{(2)}_{\rm f}\!\!\!:\!\!|-|\!\!:\!\!n^{(1)}_{\rm f}\!\!\!:\!\!|}\,.
\eeq
Assuming for simplicity and from now on that $N_{\rm s}$ is divisible for both $|\!\!:\!\!n^{(1)}_{\rm f}\!\!\!:\!\!|$ and $|\!\!:\!\!n^{(2)}_{\rm f}\!\!\!:\!\!|$, let us start by taking $|\!\!:\!\!n^{(1)}_{\rm f}\!\!\!:\!\!|=1$ and $|\!\!:\!\!n^{(2)}_{\rm f}\!\!\!:\!\!|=2$. In this case, we have that $\mu=2E(N_{\rm s}/2)-E(N_{\rm s})$. Next, assuming $\delta_{x}\equiv \delta=const$, we can express $E(N_{\rm s}/2)=E(N_{\rm s})-(N_{\rm s}/2)\delta$, so $\mu=E(N_{\rm s})-N_{\rm s}\delta$. If we repeat the calculations for $|\!\!:\!\!n^{(2)}_{\rm f}\!\!:\!\!|=3$, we would get that $E(N_{\rm s}/3)=E(N_{\rm s})-(2N_{\rm s}/3)\delta$ and $\mu'=[3E(N_{\rm s}/3)-3(2N_{\rm s}/3)\delta-E(N_{\rm s})]/2=E(N_{\rm s})-N_{\rm s}\delta = \mu$. Indeed, it is not hard to see that in the $\delta_{x}=const$-limit the chemical potential is the same for an arbitrary $|\!\!:\!\!n^{(2)}_{\rm f}\!\!:\!\!|$, what explains the transition from half filling $\rho=0$ to the saturated regime $\rho=1/a$ at $g^2=0$. As we increase interactions, $\delta_{x}$ becomes slightly less flat, and the system transitions many different fillings when increasing $\mu$, forming the compressible phase shown in Fig.~\ref{fig:rho_kappa_gsq}.

The characteristic plateaus of the incompressible phases, which appear at fillings $\rho=(|\!\!:\!\!n_{\rm f_0}\!\!:\!\!|\,a)^{-1}$ for small integers $|\!\!:\!\!n_{\rm f_0}\!\!:\!\!|$, can also be explained through the Hilbert-space fragmentation. To illustrate that, let us assume a $|\!\!:\!\!n^{(1)}_{\rm f}\!\!\!:\!\!|$ such that $|\!\!:\!\!n^{(1)}_{\rm f}\!\!\!:\!\!|/N_{\rm s}=1/n$ and a $|\!\!:\!\!n^{(2)}_{\rm f}\!\!\!:\!\!|$ fulfilling $|\!\!:\!\!n^{(2)}_{\rm f}\!\!\!:\!\!|/N_{\rm s}=1/(n-1)$. Following the same ideas as before, one can check that for these two fillings $\mu=E(n)-|\!\!:\!\!n^{(2)}_{\rm f}\!\!\!:\!\!|\delta_{n-1}/(|\!\!:\!\!n^{(2)}_{\rm f}\!\!\!:\!\!|-|\!\!:\!\!n^{(1)}_{\rm f}\!\!\!:\!\!|)=E(n)-n\delta_{n-1}$. Next, let us take another filling $|\!\!:\!\!n^{(3)}_{\rm f}\!\!\!:\!\!|$ such that $|\!\!:\!\!n^{(1)}_{\rm f}\!\!\!:\!\!|<|\!\!:\!\!n^{(3)}_{\rm f}\!\!\!:\!\!|<|\!\!:\!\!n^{(2)}_{\rm f}\!\!\!:\!\!|$, a condition mainly fulfilled for several $|\!\!:\!\!n^{(3)}_{\rm f}\!\!\!:\!\!|$ if $n$ is small enough compared with $N_{\rm s}$. Repeating an analogous procedure as before, it can be shown that, since it is not necessary to use the curvature of $\delta_{x}$ for the calculations, but only its evaluation at $n-1$, we recover in practice a similar scenario as for $\delta_{x}=const$, obtaining that $\mu'=E(n) -n\delta_{n-1}=\mu$, what implies that system goes directly from $|\!\!:\!\!n^{(1)}_{\rm f}\!\!\!:\!\!|$ to $|\!\!:\!\!n^{(2)}_{\rm f}\!\!\!:\!\!|$. This, together with the fact that $\mu$ increases as we reduce $n$, explains the plateaus.

\subsection{Chiral spirals by lifting the fragmentation}\label{sec:chiral_spirals}

\begin{figure}
    \centering
    \includegraphics[width=0.9\linewidth]{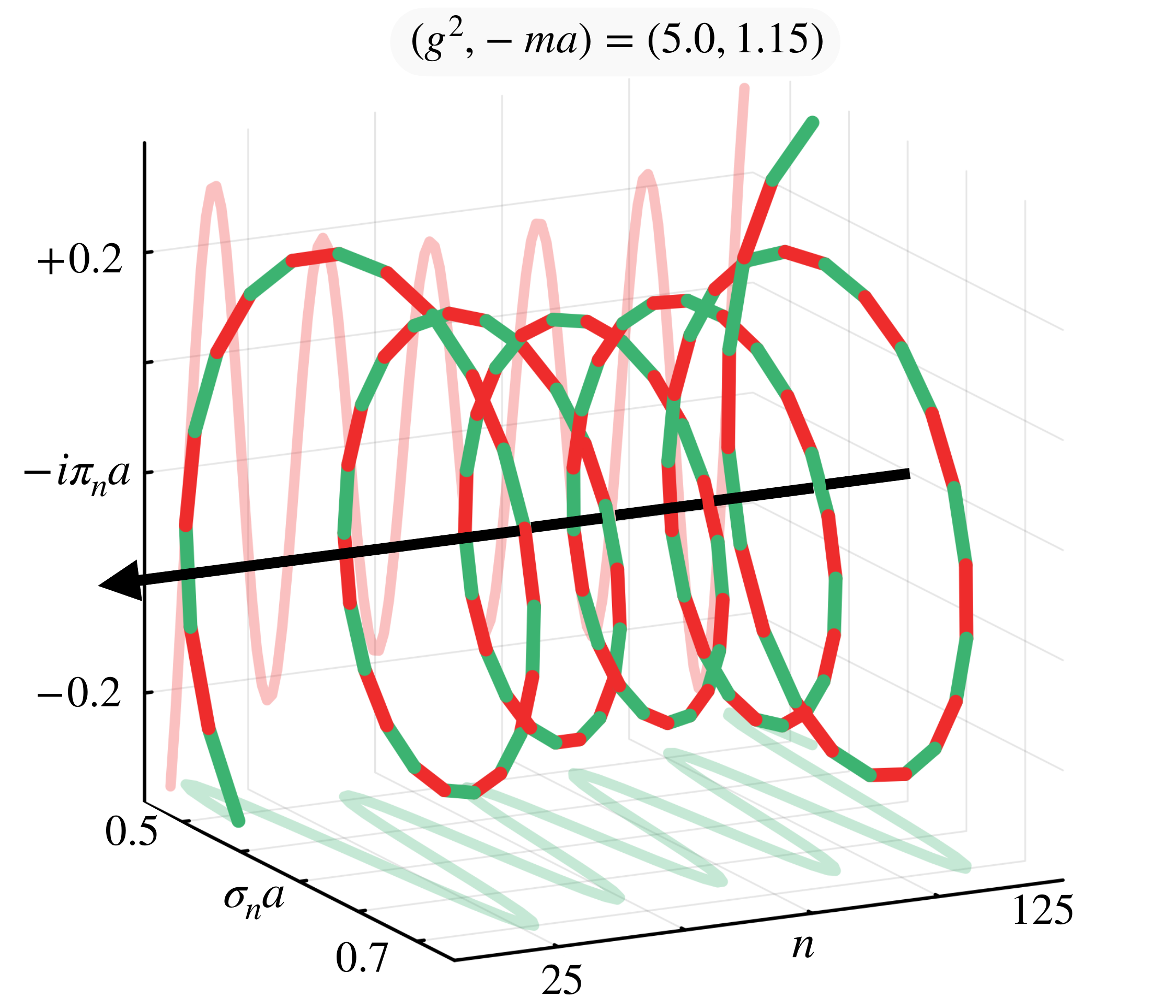}
    \caption{\textbf{Complex condensate $\boldsymbol{\Delta_n}$ at $\mathbf{(g^2, -ma)=(5.0,\,1.15)}$ and $\mathbf{:\!n_{\rm f}\!:=+6}$.} As seen in Fig.~\ref{fig:inhomogeneities_out_of_ma-1}, far away from the symmetry line but inside the zero-density parity broken phase both condensates behave as a sine/cosine with a wavevector $k=2\pi\rho$ in the bulk, so that the complex condensate $\Delta$ corresponds to an elliptical plane wave.}
    \label{fig:chiral_spiral_3D}
\end{figure}

\begin{figure*}
    \centering
    \includegraphics[width=1.0\linewidth]{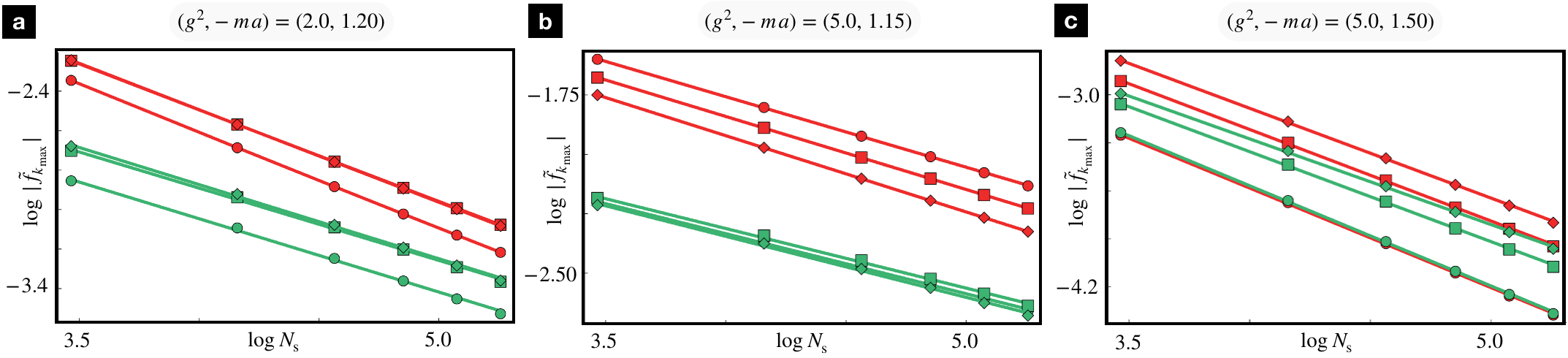}
    \caption{\textbf{Oscillation scaling out of the symmetry line $\boldsymbol{ma=-1}$ for several densities $\boldsymbol{\rho}$.} Scaling of the maximum amplitude spectrum $|\tilde{f}_{k_{\rm max}}|$ of the pseudoscalar (red) and scalar (green) condensates with the number of sites $N_{\rm s}$, with a previous subtraction of their spatial mean --denoted with a tilde. This analysis has been made for the densities $\rho a\!=\!0.125$ (circle) --commensurate--, $\rho a\!=\!0.1875$ (square) --incommensurate-- and $\rho a\!=\!0.25$ (diamond) --commensurate--, at three representative points: \textbf{(a)} $(g^2,\, -ma)=(2.0,\,1.20)$, \textbf{(b)} $(g^2,\,-ma)=(5.0,\,1.15)$, and \textbf{(c)} $(g^2,\,-ma)=(5.0,\,1.50)$, which are inside the half-filled SPT, Aoki and trivial phases, respectively. As can be seen, in all cases the spectrum amplitude decays to zero in the thermodynamic limit following a power law $|\tilde{f}_{k_{\rm max}}|=f_0\,N_s^{-\alpha}+f_{\infty}$. Likewise, the decay exponents seem to depend mainly on $g^2$ and $ma$, presenting slight variations with the density for both condensates.}
    \label{fig:oscillation_scaling}
\end{figure*}

The data shown in Figs.~\ref{fig:inhomogeneities_out_of_ma-1}--\ref{fig:chiral_spiral_3D} confirm the expectation, discussed in the introduction, that the parity-breaking phase of the GNW model supports inhomogeneous condensates with a chiral spiral structure. This behaviour becomes manifest when moving sufficiently far from the symmetry line \( ma = -1 \), region at which the dimer particle numbers are no longer conserved charges due to the appearance of the terms $(ma+1)/a\,\,\Phi^\dagger_{n,-}\Phi_{n,+}+{\rm H.c.}$ in the GNW Hamiltonian. More specifically, in the regime \( g^2 \gtrsim 5.0 \) we observe that the pseudoscalar and scalar condensates develop coherent oscillations with fixed relative phase and a well-defined wavevector \( k = 2\pi \rho \). As seen in Fig.~\ref{fig:inhomogeneities_out_of_ma-1} panels \textbf{(a)--(b)}, the condensates transition from discontinuous anti-kinks near the symmetry line to bulk modulations that closely resemble cosine and sine waves. These oscillations are accompanied by a similarly modulated excess in the rung density, whose maxima align with regions of negative slope in \( \pi_n \), consistent with the soliton-bound charge picture established in earlier sections.

The elliptical nature of these modulations is further confirmed by reconstructing the complex condensate \( \Delta_n = \sigma_n - \ii\pi_n \), depicted in Fig.~\ref{fig:chiral_spiral_3D}. In the parity-broken phase far from \( ma = -1 \), \( \Delta_n \) traces a helical trajectory in the complex plane, forming an elliptical plane wave in the bulk, since generically both condensates have different amplitude modulations.  This structure is reminiscent of a chiral spiral: a spatially oscillating condensate where the U(1) chiral phase angle rotates uniformly along the chain. Indeed, since both condensates exhibit the same spatial modulation up to a relative phase of $\pi/2$, we can certainly think of our numerical solutions as a superposition of two chiral spirals with opposite wavevector. In any case, these results validate the picture of quasi-spiral symmetry breaking in the incommensurate regime and provide strong non-perturbative evidence for the existence of such textures in lattice-regulated relativistic field theories.

Notably, the onset of the chiral spiral regime is not universal but depends sensitively on coupling, filling, and chain length. Concerning the first two, if we maintain the coupling strength to $g^2=5$ and move further away from the $ma=-1$, out of the parity-breaking phase at zero density, the oscillations become highly suppressed, as can be seen in Fig.~\ref{fig:inhomogeneities_out_of_ma-1} \textbf{(c)}. This suppression also takes place for weaker couplings, as shown in Fig.~\ref{fig:inhomogeneities_out_of_ma-1} \textbf{(d)}, even moving slightly away from \( ma = -1 \), with the difference that in this case the condensates retain residual features from the zero-density topological edge states. This suggests that the spiral condensate depends on a combination of finite density, sufficiently strong interactions, and the absence of competing commensurate crystalline order. Altogether, the emergence of the chiral spiral represents the continuum limit of the soliton lattice picture developed at low densities, completing the narrative arc from fragmented crystals to smooth, delocalised topological textures.

Aside from the coupling dependence, we observe that the modulations decay smoothly from the boundaries. Through a more careful inspection, we show in Fig.~\ref{fig:oscillation_scaling} that these modulations --suppressed or not by the couplings-- generically decay to zero following a power law. More specifically, we have computed the Discrete Fourier Transform of both condensates, and retained the peak amplitude spectrum $|\tilde{f}_{k_{\rm max}}|\equiv |\tilde{\pi}_{k_{\rm max}}|,\,|\tilde{\sigma}_{k_{\rm max}}|$ --subtracting the mean value along the chain-- for several number of sites $N_s=\{32,\,64,\,96,\,128,\,160,\,192\}$. As a numerical remark, in order to capture correctly the correlations, we have repeated the simulations increasing the bond dimension until convergence for the condensates. Next, we have fitted these maximum amplitude spectra to $|\tilde{f}_{k_{\rm max}}|=f_0\,N_s^\alpha+f_{\infty}$ for several commensurate and incommensurate fillings at different points of the phase diagram, corresponding to the various phases at zero density as a probe to search for distinct behaviours. In particular, we have taken $(g^2,\, -ma)=(2.0,\,1.20)$ as representative of the SPT phase, $(g^2,\,-ma)=(5.0,\,1.15)$ for the Aoki phase, and $(g^2,\,-ma)=(5.0,\,1.50)$ for the trivial one. In all cases, the power law decay fits with good precision. Altogether, we conclude that, in the absence of an exhaustive exploration of the whole phase diagram, the chiral spirals do not present long-range modulated order.

\section{\bf Conclusions and Outlook}\label{sec:5}

In this work, we have presented a detailed study of the Gross-Neveu-Wilson model at finite fermion density in the single-flavour limit \( N=1 \). Using MPS simulations in both grand-canonical and canonical ensembles, we have uncovered a rich structure of ground states beyond the conventional homogeneous condensates described by large-\( N \) methods. Our results demonstrate the existence of inhomogeneous condensates, crystalline charge distributions, and quasi-chiral spirals.

Along the symmetry line \( ma = -1 \), we have shown that the model exhibits Hilbert-space fragmentation, leading to the formation of crystals composed of immobile interfaces that confine the doped fermions or holes, for arbitrary interactions. If these are weak, we observe topological crystals with periodically-distributed topological defects hosting one particle each. Conversely, for stronger interactions, the system enters into a parity-broken phase where the doped fermions induce solitonic distortions in the pseudoscalar condensate, forming anti-kinks with quantised topological charge. As the density increases, these kinks organise into regular soliton lattices. Finally, by changing the value of the bare mass, we find that the inhomogeneous patterns culminate in smooth periodic, antiphase modulations for the scalar and pseudoscalar condensates, which reminds one of chiral spirals, with the difference that these are elliptical and seem to show a power-law decay.

These results offer a nonperturbative confirmation of long-suspected inhomogeneous phases in 1D Gross-Neveu-type models with discrete $\mathbb{Z}_2$ chiral symmetry, unveiling a new microscopic mechanism based on fragmentation. Beyond their theoretical interest, our findings are of direct experimental relevance. The GNW model can be realised in cold-atom quantum simulators using Raman-induced optical ladders with tunable Hubbard interactions~\cite{doi:10.1142/9789813272538_0001,PhysRevLett.121.150401,doi:10.1126/sciadv.aao4748,doi:10.1126/science.aaf6689,PhysRevResearch.5.L012006}, and the inhomogeneous textures reported here may be probed via quantum gas microscopy. In this context, the solitonic and crystalline patterns serve as   analogs of baryonic matter, while the chiral spiral  could emulate aspects of dense QCD matter.

In future work, it would be natural to investigate the role of temperature and a more elaborate characterisation of entanglement, distinguishing crossover from critical phenomena. Extensions to multi-component fermions (\( N>1 \)) and higher-dimensional Gross-Neveu models would also allow for a more direct analogy with QCD predictions. For the latter extension, one can use the Projected Entangled Pair States (PEPS) tensor network ansatz \cite{verstraete2004renormalization, PhysRevLett.96.220601, hyatt2019dmrg} instead of MPSs, although approximate methods to contract the network are required \cite{orus2019tensor, schuch2007computational, lubasch2014unifying, lubasch2014algorithms}. Likewise, it would be interesting to explore whether the fragmentation mechanism and soliton confinement observed here generalise to other classes of discrete or gauge-symmetric models, offering a broader platform for studying non-perturbative QFTs using tensor networks and quantum simulation. Also, generalisations  to higher-dimensional variants of this model, like for instance $(2+1)$-dimensional square lattices, could be explored. As discussed in \cite{10.21468/SciPostPhys.17.1.003}, one may add an anisotropic twisted Wilson mass in order to find a flat-band regime analogous to the one observed in $(1+1)$ dimensions for $ma=-1$, leading in this case to correlated higher-order topological phases.  When  modifying the fermion density, we  expect to have a similar phenomenon of Hilbert space fragmentation. Nevertheless, possible doping-based fragmentations in this higher-dimensional setting  will likely not lead to the fragmentation of the model into fully disconnected pieces, which simplified our analysis on the one-dimensional case. Lastly, coming back to open questions for the current model,  a next step to fully understand the inhomogeneities covered in this paper would be to scan in greater detail if there are regions of the phase diagram out of the symmetry line, i.e. possible critical lines, where the elliptic modulations become long range, and whether the amplitudes of the condensates coincide therein, to fully match with the analytic expectation.
\acknowledgments
A.B acknowledges support from PID2021- 127726NB-
I00 (MCIU/AEI/FEDER, UE), from the Grant IFT Cen-
tro de Excelencia Severo Ochoa CEX2020- 001007-S,
funded by MCIN/AEI/10.13039/501100011033, from the
CAM/FEDER Project TEC-2024/COM 84 QUITEMAD-CM,
and from the CSIC Research Platform on Quantum Tech-
nologies PTI- 001. {S.C thanks support from FPU23/02915 scholarship from the MCIU.}

\bibliography{manuscript4}
\end{document}